\documentclass[times,twocolumn,final]{elsarticle}
\usepackage{framed, multirow}
\usepackage{soul,color}

\usepackage{amssymb}
\usepackage{latexsym}

\usepackage{url}
\usepackage[dvipsnames]{xcolor}

\usepackage{hyperref}

\definecolor{newcolor}{rgb}{.8,.349,.1}

\usepackage{lipsum}
\usepackage{float}
\usepackage{caption}
\usepackage{booktabs}	% cool tables
\usepackage{longtable}
\usepackage{array}
\usepackage{multirow}
\usepackage{threeparttable}	% table footnotes!
\usepackage{url}
\usepackage{amsmath}
\usepackage{xfrac}

\newcolumntype{C}[1]{>{\centering\let\newline\\\arraybackslash\hspace{0pt}}m{#1}}
\usepackage{subfigure}

\journal{Medical Image Analysis}

\begin{document}

%\verso{A. Signoroni, M. Savardi \textit{et~al.}}

\begin{frontmatter}

%\title{End-to-end learning for semiquantitative rating of COVID-19 severity on Chest X-rays}
\title{BS-Net: learning COVID-19 pneumonia severity on a large Chest X-Ray dataset}
%\title{Learning COVID-19 pneumonia severity scores from Chest X-rays}
%\title{Learning severity scores for Chest X-rays COVID-19 pneumonia}

%\title{End-to-end learning for semiquantitative rating of COVID-19 severity on Chest X-rays in the heart of southern Europe outbreak}
%\title{End-to-end learning of COVID-19 severity assessment on 5000 Chest X-Rays of hospitalized patients in northern Italy}

%Mattia: End-to-end learning for COVID-19 automatic severity estimation on Chest X-rays

\author[unibs-dii]{Alberto Signoroni\corref{cor1}\fnref{fn1}}
\ead{alberto.signoroni@unibs.it}
\author[unibs-dii]{Mattia Savardi\fnref{fn1}}
\author[unibs-dii]{Sergio Benini}
\author[unibs-dii]{Nicola Adami}
\author[unibs-dii]{Riccardo Leonardi}
\author[unibs-dsmc]{Paolo Gibellini}
\author[unibs-dsmc]{Filippo Vaccher}
\author[unibs-dsmc]{Marco Ravanelli}
\author[unibs-dsmc]{Andrea Borghesi}
\author[unibs-dsmc]{Roberto Maroldi}
\author[unibs-dsmc]{Davide Farina}

\address[unibs-dii]{Department of Information Engineering -- University of Brescia, Brescia, Italy}
\address[unibs-dsmc]{Department of Medical and Surgical Specialties, Radiological Sciences, and Public Health -- University of Brescia, Brescia, Italy}

\cortext[cor1]{Corresponding author}
\fntext[fn1]{Shared first authorship}

%\accepted{\textit{Accepted on Medical Image Analysis, 18-03-2021}}

\begin{abstract}
In this work we design an end-to-end deep learning architecture for predicting, on Chest X-rays images (CXR), a multi-regional score conveying the degree of lung compromise in COVID-19 patients.
Such semi-quantitative scoring system, namely \texttt{Brixia~score}, is applied in serial monitoring of such patients, showing significant prognostic value, in one of the hospitals that experienced one of the highest pandemic peaks in Italy.
To solve such a challenging visual task, we adopt a weakly supervised learning strategy structured to handle different tasks (segmentation, spatial alignment, and score estimation) trained with a ``from-the-part-to-the-whole" procedure involving different datasets. In particular, we exploit a clinical dataset of almost 5,000 CXR annotated images collected in the same hospital. Our BS-Net demonstrates self-attentive behavior and a high degree of accuracy in all processing stages. Through inter-rater agreement tests and a gold standard comparison, we show that our solution outperforms single human annotators in rating accuracy and consistency, thus supporting the possibility of using this tool in contexts of computer-assisted monitoring. Highly resolved (super-pixel level) explainability maps are also generated, with an original technique, to visually help the understanding of the network activity on the lung areas.
We also consider other scores proposed in literature and provide a comparison with a recently proposed non-specific approach.
We eventually test the performance robustness of our model on an assorted public COVID-19 dataset, for which we also provide \texttt{Brixia~score} annotations, observing good direct generalization and fine-tuning capabilities that highlight the portability of BS-Net in other clinical settings.
The CXR dataset along with the source code and the trained model are publicly released for research purposes.
\end{abstract}


\end{frontmatter}

%\linenumbers

%% -----------------------------------------
%%          SECTION 1 - INTRODUCTION
%% -----------------------------------------

%% main text
\section{Introduction}
\label{sec:intro}

% 1-Outbrake and saturation of facilities
%In Italy, starting February 2020, Lombardy region was the epicenter of the outbreak of COVID-19, and Brescia was among the most severely involved provinces. At the peak of the epidemic, the city university hospital (ASST Spedali Civili di Brescia) admitted up to 900 COVID-19 patients.

Worldwide, the saturation of healthcare facilities, due to the high contagiousness of Sars-Cov-2 virus and the significant rate of respiratory complications \citep{who2020}, is indeed one among the most critical aspects of the ongoing COVID-19 pandemic. Under these conditions, it is extremely important to adopt all types of measures to improve the accuracy in monitoring the evolution of the disease and the level of coordination and communication between different clinicians for the streamlining of healthcare procedures, from facility- to single patient-level.
\begin{figure*}[t]
    \centering
    \includegraphics[width=\textwidth]{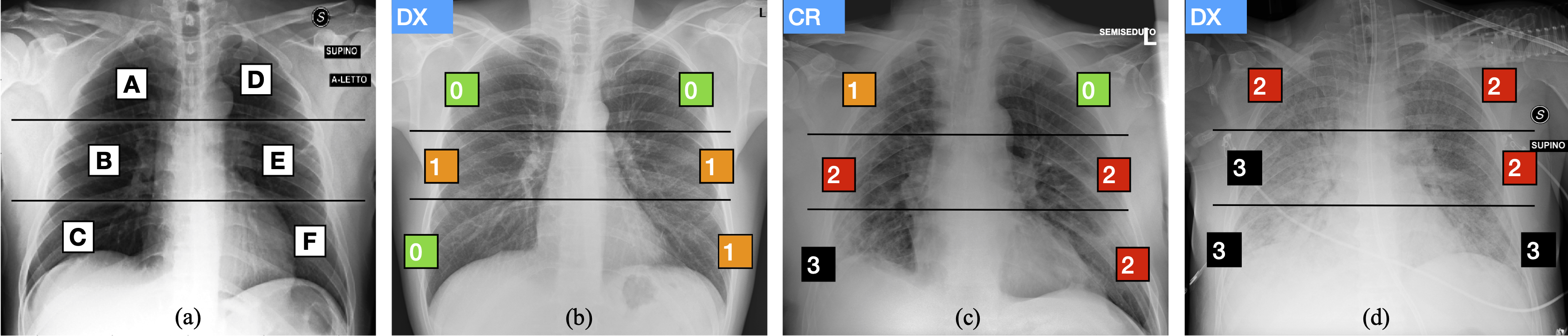}
    \caption{\texttt{Brixia~score}: (a) zone definition and (b-c-d) examples of annotations. Lungs are first divided into six zones on frontal chest x-rays. Line A is drawn at the level of the inferior wall of the aortic arch. Line B is drawn at the level of the inferior wall of the right inferior pulmonary vein. A and D upper zones; B and E middle zones; C and F lower zones. A score ranging from 0 (green) to 3 (black) is then assigned to each sector, based on the observed lung abnormalities.}
    \label{fig:bscore}
\end{figure*}
% 2-why CXR
In this context, thoracic imaging, specifically chest X-ray (CXR) and computed tomography (CT), is playing an essential role in the management of patients, especially those evidencing risk factors (from triage phases) or moderate to severe COVID-19 signs of pulmonary disease \citep{rubin2020}. In particular, CXR is a widespread, relatively cheap, fast, and accessible diagnostic modality, which may be easily brought to the patient's bed, even in the emergency departments. Therefore, the use of CXR may be preferred to CT not only in limited resources environments, but where this becomes fundamental to avoid handling of more compromised patients, or to prevent infection spread due to patient movements from and towards radiology facilities \citep{manna2020}. Moreover, a serious X-ray dose concern arises in this context: due to the typical rapid worsening of the disease and the need for a prompt assessment of the therapeutic effects, serial image acquisitions for the same patient are often needed on a daily basis.
For these reasons, notwithstanding a lower sensitivity compared to CT, CXR has been set up as a first diagnostic imaging option for COVID-19 severity assessment and disease monitoring in many healthcare facilities.
% 3-Severity score
%Il merito dell'ASST è stato quello di intrrodurre un scoree particolarmente espressiivo sin dalle priime fasi della pandemia.

Due to the projective nature of the CXR image and the wide range of possible disease manifestations, a precise visual assessment of the entity and severity of the pulmonary involvement is particularly challenging. To this purpose scoring systems have been recently adopted to map radiologist judgments to numerical scales, leading to a more objective reporting and improved communication among specialists. In particular, from the beginning of the pandemic phase in Italy, the Radiology Unit 2 of ASST Spedali Civili di Brescia introduced a multi-valued scoring system, namely \texttt{Brixia~score}, that was immediately implemented in the clinical routine \citep{borghesi2020first}. With this system, the lungs are divided into six regions, and the referring radiologist assigns to each region an integer rating from 0 to 3, based on the local assessed severity of lung compromise (see Figure~\ref{fig:bscore} and Section~\ref{ssec:Brixia} for details). 
Other scores have been recently proposed as well, which are less resolute either in terms of intensity scale granularity \citep{toussie2020}, or localization capacity \citep{wong2020,cohen2020predicting} (see Section~\ref{ssec:OtherSeverity}). 
% 5- AI solution for computer-aided interpretation of covid-19 images. From diagnosis to severity assessment
Severity scores offer a mean to introduce the use of representation learning techniques to automatically assess disease severity using artificial intelligence (AI) approaches starting from CXR analysis.
However, while in general computer-aided interpretation of radiological images based on Deep Learning (DL) can be an important asset for a more effective handling of the pandemic in many directions \citep{shi2020reviewAI-Covid}, the early availability of public datasets of CXR images from COVID-19 subjects catalyzed the research almost exclusively on assisted COVID-19 diagnosis (i.e., dichotomous or differential diagnosis versus other bacterial/viral forms of pneumonia).
This happened despite the fact that normal CXRs do not exclude the possibility of COVID-19, and abnormal CXR is not enough specific for a reliable diagnosis \citep{rubin2020}. %eventualmente aggiungi altre ref
Moreover, much research applying AI to CXR in the context of COVID-19 diagnosis considered small or private datasets, or lacked rigorous experimental methods, potentially leading to overfitting and performance overestimation \citep{castiglioni2020, maguolo2020critic, tartaglione2020unveiling}.
So far, no much work has been done in other directions, such as COVID-19 severity assessment from CXR, despite this has been highlighted as one of the highest reasonable research efforts to be pursued in the field of AI-driven COVID-19 radiology \citep{laghi2020e225, cohen2020data}. %verifica su Laghi se proprio così
\begin{figure*}[t]
   \centering
    \includegraphics[width=\textwidth, height=8cm]{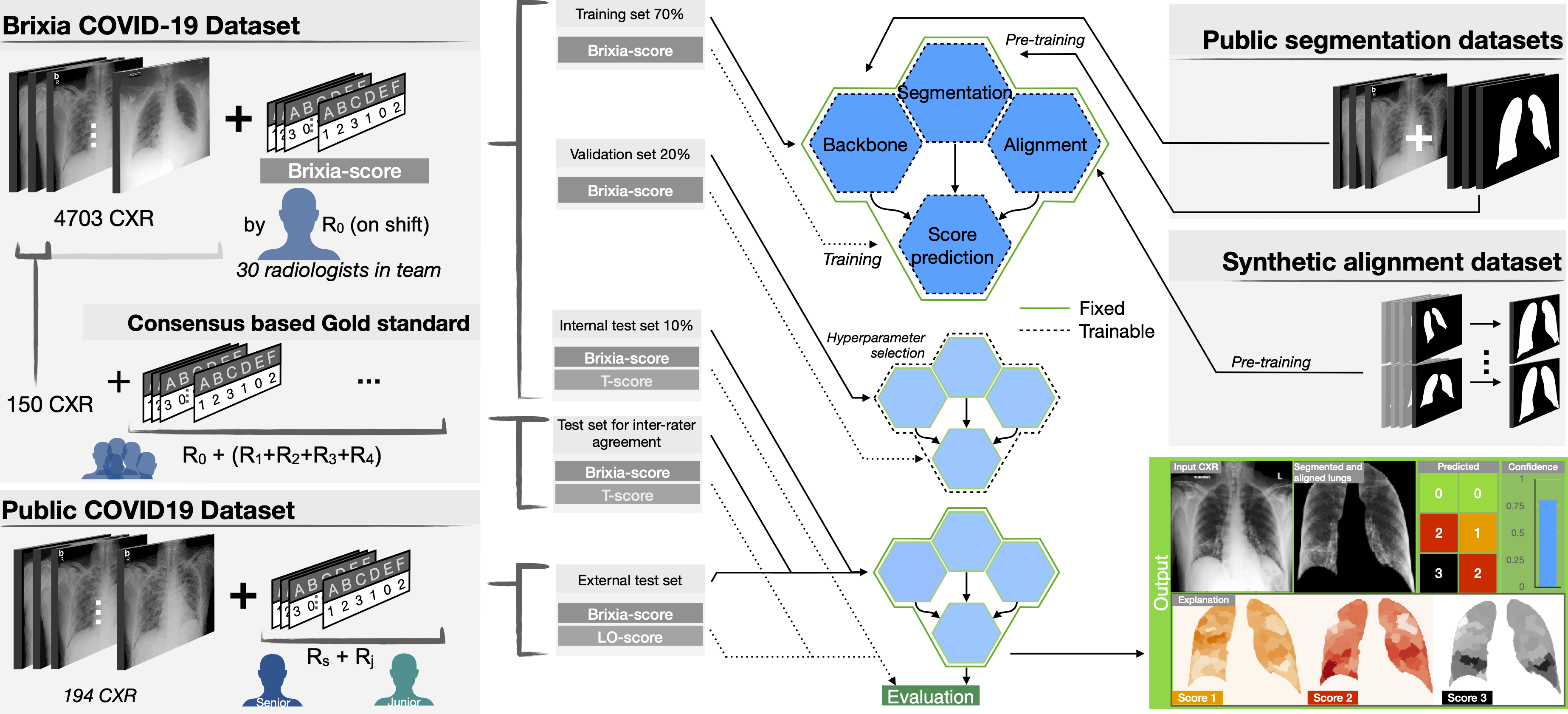}
    \caption{Overview of the proposed method: representation of the two COVID-19 datasets (on the left) with associated \textit{Brixia score} annotations, and of the other two datasets (on the right) used for the pre-training. Datasets splitting and usage is indicated (in the middle) for training/validation/test phases. The outputs of the proposed system are illustrated as well (bottom right). }
    \label{fig:framework}
\end{figure*}

\subsection {Aims and contributions}
\label{ssec:aims}

\textit{Main goal.} In this work, we aim at introducing the first learning-based solution specifically designed to obtain an effective and reliable assessment of the severity of COVID-19 disease by means of an automated CXR interpretation. This system is able to produce a robust multi-regional self-attentive scoring estimation on clinical data directly coming from different X-ray modalities (computed radiography, CR, and digital radiography by Direct X-ray detection, DX), acquisition directions (anteroposterior, AP, and posteroanterior, PA) and patient conditions (e.g., standing, supine, with or w/o the presence of life support systems).

\textit{Large CXR database.} For this purpose, we collected and operated on a large dataset of almost 5,000 CXRs, which can be assumed representative of all possible manifestations and degrees of severity of the COVID-19 pneumonia, since these images come from the whole flow of hospitalized patients in one of the biggest healthcare facilities in northern-Italy during the first pandemic peak. 
This CXR dataset is fully annotated with scores directly coming from the patients' medical records, as provided by the reporting radiologist on duty among the about 30 specialists forming the radiology staff. Therefore, we had the possibility to work on a complete and faithful picture of an intense and emergency-driven clinical activity.

\textit{End-to-end severity assessment architecture.} We developed an original multi-block Deep Learning-based architecture designed to learn and execute different tasks at once such as: data normalization, lung segmentation, %geometric alignment, feature extraction
feature alignment, and the final multi-valued score estimation. Although the presence of specific task dedicated blocks, the overall logic is \textit{end-to-end} in the sense that the gradient flows from the head (\texttt{brixia score} estimation), to the bottom (input CXR) without interruption. This is reached through a semantically hierarchical organization of the training, of which we give a high-level representation in Figure \ref{fig:framework}. To squeeze all the information from the data, we organize the training according to a \textit{from-the-part-to-the-whole} strategy. This consists of leveraging multiple CXR datasets (described in Section~\ref{sec:data}) for early training stages, then arriving at a global (involving all network portions) severity assessment training, based on the large collected dataset. Our end-to-end network architecture, called BS-Net and described in detail in Section~\ref{sec:methods}, is thus characterized by the joint work of different sections, represented as a group of four hexagons in Figure \ref{fig:framework}, comprising 1) a shared multi-task feature extraction backbone, 2) a state-of-art lung segmentation branch, 3) an original registration mechanism that acts as a "multi-resolution feature alignment" block operating on the encoding backbone
%mechanism capable of spatially normalizing the feature computation stream in the backbone \colorbox{green}{R1.1} 
, and 4) a multi-regional classification part for the final six-valued score estimation. All these blocks act together in the final training thanks to a loss specifically crated for this task.
This loss guarantees also performance robustness, comprising a differentiable version of the target discrete metric.

\textit{Weakly supervised learning framework.} 
The learning phase operates in a weakly-supervised fashion. This is due to the fact that difficulties and pitfalls in the visual interpretation of the disease signs on CXRs (spanning from subtle findings to heavy lung impairment), and the lack of detailed localization information, produces unavoidable inter-rater variability among radiologists in assigning scores \citep{zhou2017}.
Far from constituting in itself a ``weakness", these approaches have demonstrated to be highly valuable methods to leverage available knowledge in medical domains \citep{xu2014,chestxray8,karimi2019deep,bontempi2020,tajbakhsh2020}.

\textit{Explainability maps.} In the perspective of a responsible and transparent exploitation of the proposed solution, there is the need to establish a communication channel between the specialist and the AI system. Given the spatial distribution of the severity assessment, we need highly resolved explainability maps, also able to compensate for some limitations of conventional approaches based on Grad-CAM \citep{selvaraju2017}. To this end we propose an original technique (described in Section~\ref{sec:explain}) able to create highly structured explanation maps at a super-pixel level.

\textit{Experimental goals.} Experiments presented in Section~\ref{sec:results} are carried out as follows. 1) In terms of system performance assessment, we select the best configurations and justify architectural choices also considering different possible alternatives from the literature. 2) To cope with the inter-rater variability and to demonstrate above radiologists' performance, we involve a team of specialists to establish a consensus-based gold-standard as a reference for both single radiologist and our model ratings. 3) We consider other scores proposed in the literature and provide a comparison with a recently proposed non-specific approach for demonstrating both system versatility and performance. 4) We address the problem of portability of the proposed solution on CXRs coming from different worldwide contexts by providing new annotations and assessing our system behavior on a public CXR dataset of reference for COVID-19.

\textit{Data, Code, and Model distribution.} The whole dataset along with the source code and the trained model are available from \url{http://github.com/BrixIA}.

%Beyond increasing the scope and value of the public CXR dataset, by providing additional annotations for tasks different than diagnosis, we will make use of the annotated COVID-19 repository to demonstrate the robustness and the portability of the proposed solution, even on the complex case mix of CXR present in this collection (different provenance, variable image quality,...).

%% -----------------------------------------
%% SECTION 2 - RELATED WORK
%% -----------------------------------------

\section{Related Work}
\label{sec:prior}

%%% FROM THE INTRO 4-AI solutions for pandemic
Since the very beginning of the pandemic, the attention and resources of researchers in digital technologies \citep{ting2020}, AI and data science \citep{latif2020} have been captured by COVID-19.
%%%
A review of artificial intelligence techniques in imaging data acquisition, segmentation, and diagnosis for COVID-19 has been made in \cite{shi2020reviewAI-Covid}, where authors structured previous work according to different tasks, e.g., contactless imaging workflow, image segmentation, disease detection, radiomic feature extraction, etc.

The use of convolutional neural networks (CNN) to analyze CXRs for presumptive early diagnosis and better patient handling based on signs of pneumonia, has been proposed by \citep{oh2020deep} at the very beginning of the outbreak, when a systematic collection of a large CXR dataset for deep neural network training was still problematic.
As the epidemic was spreading, the increasing availability of CXR datasets from patients diagnosed with COVID-19 has polarized almost all the research efforts on diagnosis-oriented image interpretation studies.
%In the period from late February to mid-May 2020 we found more than 60 works focusing in this direction (most of them in a pre-print format). 
As it is hard to mention them all (there are well over 100 studies, most of them in a pre-print format), the reader can refer to an early reviewing effort \cite{shi2020reviewAI-Covid}, while here we mention the ones most related to our work and derive most relevant emerging issues.
Most presented methods exploit available public COVID-19 CXR datasets \citep{kalkreuth2020covid19}.
Constantly updated collections of COVID-19 CXR images are curated by the authors of \citep{wang2020covidnet}\footnote{https://github.com/lindawangg/COVID-Net/blob/master/docs/COVIDx.md} and \citep{cohen2020covid}\footnote{https://github.com/ieee8023/covid-chestxray-dataset}.
Prior to COVID-19, large CXR datasets have been released and also used in the context of open challenges for the analysis of different types of pneumonia \cite{chestxray8}\footnote{https://www.kaggle.com/c/rsna-pneumonia-detection-challenge/data} or pulmonary and cardiac (cardiomegaly) diseases \cite{irvin2019chexpert}\footnote{https://stanfordmlgroup.github.io/competitions/chexpert/}. These datasets are usually exploited as well in works related to COVID-19 diagnosis, as in \citep{min20}, to pre-train existing networks and fine-tune them on a reduced set of COVID-19 CXRs, or to complete case categories when the focus is on differential diagnosis to distinguish from other types of pneumonia \citep{wang2020covidnet,oh2020deep,li2020covidmobilexpert}.  
However, beyond the fact that CXR modality should be more appropriate for patient monitoring and severity assessment than for primary COVID-19 detection, other issues severely affect the previous studies which employed CXR for COVID-19 diagnosis purposes \citep{burlacu2020}. In particular, many of these data-driven works are interesting in principle, but almost all would require and benefit from extensions and validation on a higher number of CXRs from COVID-19 subjects. Working with datasets of a few hundred images, when analyzed with deep architectures, often results in severe overfitting and can encounter issues generated by unbalanced classes when larger datasets are used to represent other kinds of pneumonia \citep{pereira2020covid19}. 
%Another relevant aspect is the comparison of AI performance with those of independent radiologists \citep{murphy2020multireader}.
Moreover, most of the studies issued in this emergency period are often based on sub-optimal experimental designs, and the numerous still unknowns factors about COVID-19 severely undermine the external validity and generalizability of the performance of diagnostic tests \citep{sardanelli2020}. In \citep{maguolo2020critic} it is pointed out that many CXR based systems seem to learn to recognize more the characteristics of the dataset rather than those related to COVID-19. 
This effect can be overcome, or at least mitigated, by working on more homogeneous and larger datasets, as in \citep{castiglioni2020}, or by preprocessing the data as in \citep{pereira2020covid19}.
In addition, automated lung segmentation can play an essential role in the design of a robust respiratory disease interpretation system, either diagnosis-oriented \citep{tartaglione2020unveiling} or, as we see in this work, for pneumonia severity assessment. Lung segmentation on CXR has been recently witnessing a convergence towards encoder-decoder architectures, such as in the U-Net framework \citep{ronneberger2015u}, and in the fully convolutional approach found in \cite{long2015fully}.
All these approaches share skip connections as a common key element, originally found in ResNet \citep{he2016deep} and DenseNet architectures \citep{huang2017densely}.
The idea behind skip connections is to combine coarse-to-fine activation maps from the encoder within the decoding flow.
In doing so, these models are capable of efficiently using the extracted information from different abstraction layers.
In \cite{zhou2018}, a nested U-Net is presented, bringing the idea of skip connection to its extreme. 
This approach is well suited to capture fine-grained details from the encoder network, and exploding them in the decoding branch.
Last, in critical sectors like healthcare, the lack of understanding of complex machine-learned models is hugely problematic.
Therefore, explainable AI approaches able to reveal to physicians where the model attention is directed are always desirable.
However, while Grad-CAM \citep{selvaraju2017} and similar methods can work well on diagnostic tasks (see e.g.,  \cite{karim2020deepcovidexplainer,oh2020deep,rajaraman2020iteratively,reyes2020,hryniewska2020}), this kind of approach is not enough informative to explain severity estimations, as it usually produces defocused heatmaps that hardly reveal fine details. Hece the need to find solutions to produce denser, more insightful, and more spatially detailed visual feedback to the clinician, keeping in mind perspectives of trustable deployment.

Despite some above evidenced issues, there is still an abundant ongoing effort toward AI-driven approaches for COVID-19 detection based on CT or CXR analysis, and this produced what has been recently dubbed as a deluge of articles in \cite{summers2020}, where the need to move beyond opacity detection has been remarked.
Cautions about a radiologic diagnosis of COVID-19 infection driven by deep learning have been also expressed by \cite{laghi2020e225}, who states that a more interesting application of AI in COVID-19 infection is to provide for a more objective quantification of the disease, in order to allow the monitoring of the prognostic factors (i.e., lung compromising severity) for appropriate and timely patient treatment, especially in critical conditions like the ones characterizing the management of health facilities overload. In the global race to contain and treat COVID-19, AI-based solutions have high potentials to expand the role of chest imaging beyond diagnosis, to facilitate risk stratification, disease progression monitoring, and trial of novel therapeutic targets \citep{kundu2020}. However, the numerical disproportion of research works aiming at AI-driven image-based binary COVID-19 diagnosis, as well as the diffused availability of ready-to-use (with due fine-tuning or transfer learning) DL networks, should not bias observers' conviction (as it seems instead to happen in \citep{summers2020}) toward the idea that there is no need for purposefully technology design efforts, especially for different and clinically relevant tasks, as the one tackled here. Our work shows how a dedicated technical solution, which is up to the visual difficulty of a structured severity score assessment, can lead to a significant performance and robustness boost with respect to off-the-shelf methods. Anyway, so far only a few works present AI-driven solutions for COVID-19 disease monitoring and pneumonia severity assessment based on CXR, although this modality, for the aforementioned reasons, is part of the routine practice in many institutions, like the one from which this study originates. In \citep{cohen2020predicting}, different kinds of features coming from a neural network pre-trained on non-COVID-19 CXR datasets are considered in their predictive value on the estimation of COVID-19 severity scores.
In \citep{li2020sev} a good correlation is reported between a lung based severity score judgement and machine prediction by using a transfer learning approach from a large non-COVID-19 dataset to a small COVID-19 one. However, authors acknowledge several limitations in the capability to handle image variability that can arise due to patient condition and image acquisition settings. 
Improved generalizability has been obtained in a subsequent work from same authors \citep{li2020sev2}. In \cite{zhu2020}, transfer learning vs. conventional CNN learning has been compared on a data sample derived from \cite{cohen2020covid}. In \cite{blain2020} both interstitial and alveolar opacity are classified with a modular deep learning approach (segmentation stage followed by a fine-tuned classification network). Despite the limited data sample of 65 CXRs, the study shows correlations of severity estimation with age and comorbidity factors, intensity of care and radiologists' interpretations. In \cite{amer2020covid19} a geographic extent severity score (under the form of pneumonia area over lung area ratio) was estimated and correlated with experts' judgments on 94 CXRs, using pneumonia localization and lung segmentation networks. Geographic extent and opacity severity scores were predicted in \cite{wong2020covidnets} with a modified COVID-19 detection architecture \citep{wang2020covidnet} and stratified Monte Carlo cross-validation, with measures of correlation with respect to expert annotations on 396 CXRs.
These early studies, while establishing the feasibility of a COVID-19 severity estimation on CXRs, concurrently confirm the need of a dedicated design of methods for this challenging visual task, the urgency to operate on large annotated datasets coming from real clinical settings, and the need for expressive explainability solutions.

%% -----------------------------------------
%%  SECTION 3 - BACKGROUND, AIMS AND CONTRIBUTION
%% -----------------------------------------

\section{Scoring systems for severity assessment}
\label{sec:scoring}

CXR severity scoring based on the subdivision of lungs in different regions \citep{borghesi2020first,toussie2020,wong2020} evidenced significant prognostic value when applied in serial monitoring of COVID-19 patients \citep{borghesi2020third,borghesi2020second,maroldi2020role}.
Since radiologists are asked to map a global or region-based qualitative judgment on a quantitative scale, this diagnostic image interpretation task can be defined as \textit{semi-quantitative}, i.e. characterized by a certain degree of subjectivity. 
A detailed description of three semi-quantitative scoring systems we found in use follows.

\subsection{Brixia~score}
\label{ssec:Brixia}

The multi-region and multi-valued \texttt{Brixia~score} was designed and implemented in routine reporting by the Radiology Unit 2 of ASST Spedali Civili di Brescia \citep{borghesi2020first}, and later validated for risk stratification on a large population in \citep{borghesi2020second}.
According to it, lungs in antero-posterior (AP) or postero-anterior (PA) views, are subdivided into six zones, three for each lung, as shown in Figure~\ref{fig:bscore}(a).
For each zone, a score 0 (no lung abnormalities), 1 (interstitial infiltrates), 2 (interstitial and alveolar infiltrates, interstitial dominant), or 3 (interstitial and alveolar infiltrates, alveolar dominant) is assigned, based on the characteristics and extent of lung abnormalities.
The six scores can be aggregated to obtain a Global Score in the range $[0,18]$.
%During the peak period, the \texttt{Brixia~score} has been systematically used to report CXR in COVID-19 patients. 
Examples of scores assigned to different cases are showcased in Figure~\ref{fig:bscore}(b-d).
As in daily practice CXR exams are inevitably reported by different radiologists, this combined codification of the site and type of lung lesions makes the comparison of CXR exams faster and significantly more consistent, and this allows a better handling of patients.

%\subsection{Other severity scores and possible relations among them}
\subsection{Toussie score}
\label{ssec:OtherSeverity}

%\hl{visto che prima citiamo TRE score, io farei 3 paragrafi: brixia, Toussie, GE-LO}
In \citep{toussie2020} the presence/absence of pulmonary COVID-19 alterations is mapped on a 1/0 score associated to each of six pulmonary regions according to a subdivision scheme substantially reproducing the one of the \texttt{Brixia~score}. We here refer to this score as \texttt{T~score} which globally ranges from 0 to 6. By ignoring slight differences in terms of anatomic landmarks that guide the radiologist to determine the longitudinal lung subdivisions, the \texttt{T~score} can be directly estimated from the \texttt{Brixia~score} by just mapping the set of values $\{1,2,3\}$ of the latter to the value 1 of the former. 

\subsection{GE-LO score}
In \cite{cohen2020predicting} two different COVID-19 severity scores are considered which are derived and simplified versions of a composite scoring system proposed by \cite{warren2018} int the context of lung oedema. Both scores are composed by couples of values, one for each lung: 1) a geographic extent score, here \texttt{GE~score}, in the integer range $[0,8]$ and 2) a lung opacity score, here \texttt{LO~score} in the integer range $[0,6]$. The \texttt{GE~score}, introduced in \cite{wong2020} for COVID-19 severity assessment, assigns for each lung a value depending on the extent of involvement by consolidation or ground glass opacity (0 = no involvement; $1 \leq 25\%; 2 = 25-50\%; 3 = 50-75\%; 4 \geq 75\%$ involvement). While this area-based quantification has no clear correspondence to the judgment made with the \texttt{Brixia~score}, a possible mapping can be estimated with the \texttt{LO~score} (e.g., by a simple linear regression, as we will see in  Section~\ref{res:other-scores}), which assigns for each lung a value depending on degree of opacity (0 = no opacity; 1 = ground glass opacity; 2 = consolidation; 3 = white-out). A global score derived by a modified version of the ones introduced in \cite{warren2018} has been also used in \cite{li2020sev}.

\subsection{AI-based prediction of severity scores}
\label{ssec:AI-based}

At first sight, an automatic assessment of a semi-quantitative prognostic score may seem easier than other tasks, such as differential diagnosis, or purely quantitative severity evaluations.
Nevertheless, when dealing with semi-quantitative scores, major critical aspects arise. 

First, the difficulty of establishing a ground truth information, since subjective differences in scoring are expressed by different radiologists while assessing and qualifying the presence of, sometimes subtle, abnormalities. This differs from more quantitative tasks that can be associated to measurable targets, as in the case of DL-based quantitative (volumetric) measure of opacities and consolidations on CT scans for lung involvement assessment in COVID-19 pneumonia \citep{huang2020CT,gozes2020rapid,less2020}. The application of quantitative methods to AI-driven assessment of COVID-19 severity on CXR, however, is not advisable due to the projective nature of these images, with inherent ambiguities in relating opacity area measures to corresponding volumes.

A semi-quantitative scoring system can instead leverage the sensitivity of CXR as well as the ability of radiologists to detect COVID-19 pneumonia and communicate in an effective way its severity according to an agreed severity scale. 
%\hl{una domanda che mi sono sempre fatto: anche gli altri due score (Toussie, etc.) immagino siano assegnati da umani. Ma allora sono quantitative o semi-quantitative anche loro? Perche' non quantitative e basta? Com'e' uno score quantitative puro?}

Second, the exact localization of the lung zones and of severity-related findings (even within each of the selected lung zones) remains implicit and related to the visual attention of the specialist in following anatomical landmarks (without any explicit localization information indicated with the score, nor any lung segmentation provided). 
This results in the difficulty to define reference spatial information usable as ground truth and implies significantly incomplete annotations with respect to the task.

Eventually, the same visual task related to global or localized COVID-19 severity assessment is challenging in itself, since CXR findings may be extremely variable (from no or subtle signs to extensive changes that modify anatomical patterns and borders), and the quality of the information conveyed by the images may be impaired from the presence of medical devices, or from sub-optimal patient positioning.

All these factors, if not handled, can impact in an unpredictable way on the reliability of an AI-based interpretation. 
This concomitant presence of quantitative and qualitative aspects, on a difficult visual analysis task, makes the six-valued \texttt{Brixia~score} estimation on CXR particularly challenging. 

%% -----------------------------------------
%%     SECTION 4 - DATASETS
%% -----------------------------------------

\section{Dataset}
\label{sec:data}

Training and validation of the proposed multi-network architecture (Figure \ref{fig:framework}) take advantage of the use of multiple datasets, which are described in the following.
%For training, validation, and testing phases our architecture interacts with 5 CXR datasets 
%(see Figure \ref{fig:framework}):
%\begin{itemize}
%    \item the collected dataset at ASST Spedali Civili of Brescia, including 4,707 clinical CXR from COVID-19 patients;
%    \item a public dataset of about 200 COVID-19 CXR from several centers;
%    \item a public non-COVID-19 CXR dataset annotated for lung segmentation;
%    \item and a synthetic dataset for image alignment (rotation, scale, translation).
%\end{itemize}
%All these datasets are described in the following.

\subsection{Segmentation datasets}
\label{ssec:data-seg}
For the segmentation module we exploit and merge three different datasets: Montgomery County \citep{jaeger2014two}, Shenzhen Hospital \citep{stirenko2018chest}, and JSRT databases \citep{shiraishi2000development} with the lung mask annotated by \cite{vgin2006}, for a total of about 1,000 images. 
When indicated we adopt the original training/test set splitting (as for the JSRT database); otherwise, we consider the first 50 images as test set, and the remaining as training set (see Table \ref{tab:segmentation-dataset}). 

\begin{table}[ht]
\centering
\small
\caption{Segmentation datasets.} 
\begin{tabular}{@{}llll@{}}
\toprule
                  & Training-set & Test-set & Split \\ \midrule
Montgomery County & 88           & 50       & first 50             \\
Shenzhen Hospital & 516          & 50       & first 50              \\
JSRT database     & 124          & 123      & original            \\ \bottomrule
Total             & 728          & 223      &                \\ 
\end{tabular}
\label{tab:segmentation-dataset}
\end{table}

\subsection{Alignment dataset}
\label{ssec:data-align}
CXRs acquired in a real clinical setting lack of standardized levels of magnification and alignment of the lungs.
Moreover, possible patient positions are different (standing, sitting, prone, supine) and, according to subject conditions, it is not always feasible to produce images with an ideal shooting of the chest.
To avoid the inclusion of anatomical parts not belonging to the lungs in the AI pipeline, which would increase the task complexity and introduce unwanted biases, we integrate in the network an alignment block.
This exploits the same images used for the segmentation stage to create a synthetic dataset formed by artificially transformed images (see Table \ref{tab:alignment-dataset}), including random rotations, shifts, and zooms, which are used in first phases of the training, in an on-line augmentation fashion, using the framework provided in \cite{buslaev20}.

\begin{table}[ht]
\centering
\small
\caption{Alignment dataset: synthetic transformations. The parameters refer to the implementation in Albumentation \citep{buslaev20}. In the last column is expressed the probability of application of each transformation.}
\begin{tabular}{@{}lll@{}}
\toprule
                  & Parameters (up to) & Probability \\ \midrule
Rotation & 25 degree              &     0.8           \\
Scale & 10\%          & 0.8                   \\
Shift     & 10\%           & 0.8              \\
Elastic transformation     & alpha=60, sigma=12   &    0.2           \\
Grid distortion     & step=5, limit=0.3 &     0.2         \\
Optical distortion     & distort=0.2, shift=0.05    &     0.2             \\
\bottomrule
\end{tabular}
\label{tab:alignment-dataset}
\end{table}

\subsection{Brixia COVID-19 dataset}
\label{ssec:data-brixia}

\begin{figure}
    \centering
    \includegraphics[width=0.5\textwidth]{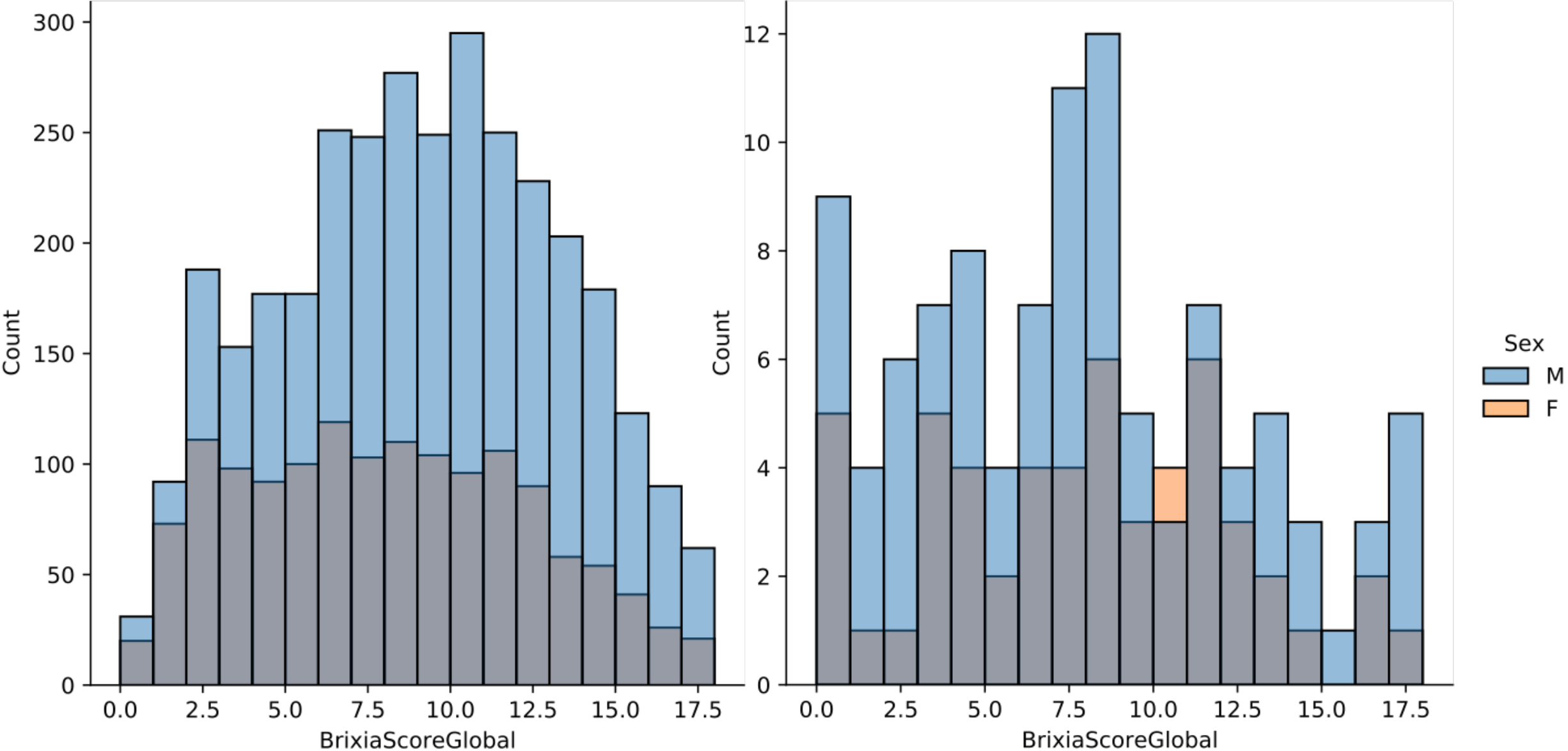}
    \caption{\texttt{Brixa score} distribution with sex stratification on the Brixia COVID-19 dataset (left), and on the dataset of \cite{cohen2020covid}~(right).}
    \label{fig:data-distr}
\end{figure}

We collected a large dataset of CXR images corresponding to the entire amount of images taken for both triage and patient monitoring in sub-intensive and intensive care units during one month (between March 4$^{th}$ and April 4$^{th}$ 2020) of pandemic peak at the ASST Spedali Civili di Brescia, and contains all the variability originating from a real clinical scenario. 
It includes 4,707 CXR images of COVID-19 subjects, acquired with both CR and DX modalities, in AP or PA projection, and retrieved from the facility RIS-PACS system.
All data are directly imported from DICOM files, consisting in 12-bit gray-scale images, and mapped to float32 between $0$ and $1$. 
To mitigate the grayscale variability in the dataset, we normalize the appearance of the CXR by sequentially applying an adaptive histogram equalization (CLAHE, clip:$0.01$), a median filtering to cope with noise (kernel size: 3), and a clipping outside the 2nd and 98th percentile.

All image reports include the \texttt{Brixia~score} as a string of six digits indicating the scores assigned to each region.
The Global Score is simply the sum of the six regional scores, and its distribution can be appreciated in Figure~\ref{fig:data-distr}~(left).
Each image has been annotated with the six-valued score by the radiologist on shift (here referred to as $R_0$), belonging to a team of about 50 radiologists operating in different radiology units of the hospital with a very wide range of years of experience and different specific expertise in imaging of the chest.
All images were collected and anonymized, and their usage for this study had the approval of the local Ethical Committee (\#$NP4121$) that also granted an authorization to release the whole anonymized dataset for research purposes.
The main characteristics of the dataset are summarised in Table~\ref{tab:dataset}.
\begin{figure*}[t!]
    \centering
    \includegraphics[width=0.95\textwidth, height=0.6\textwidth]{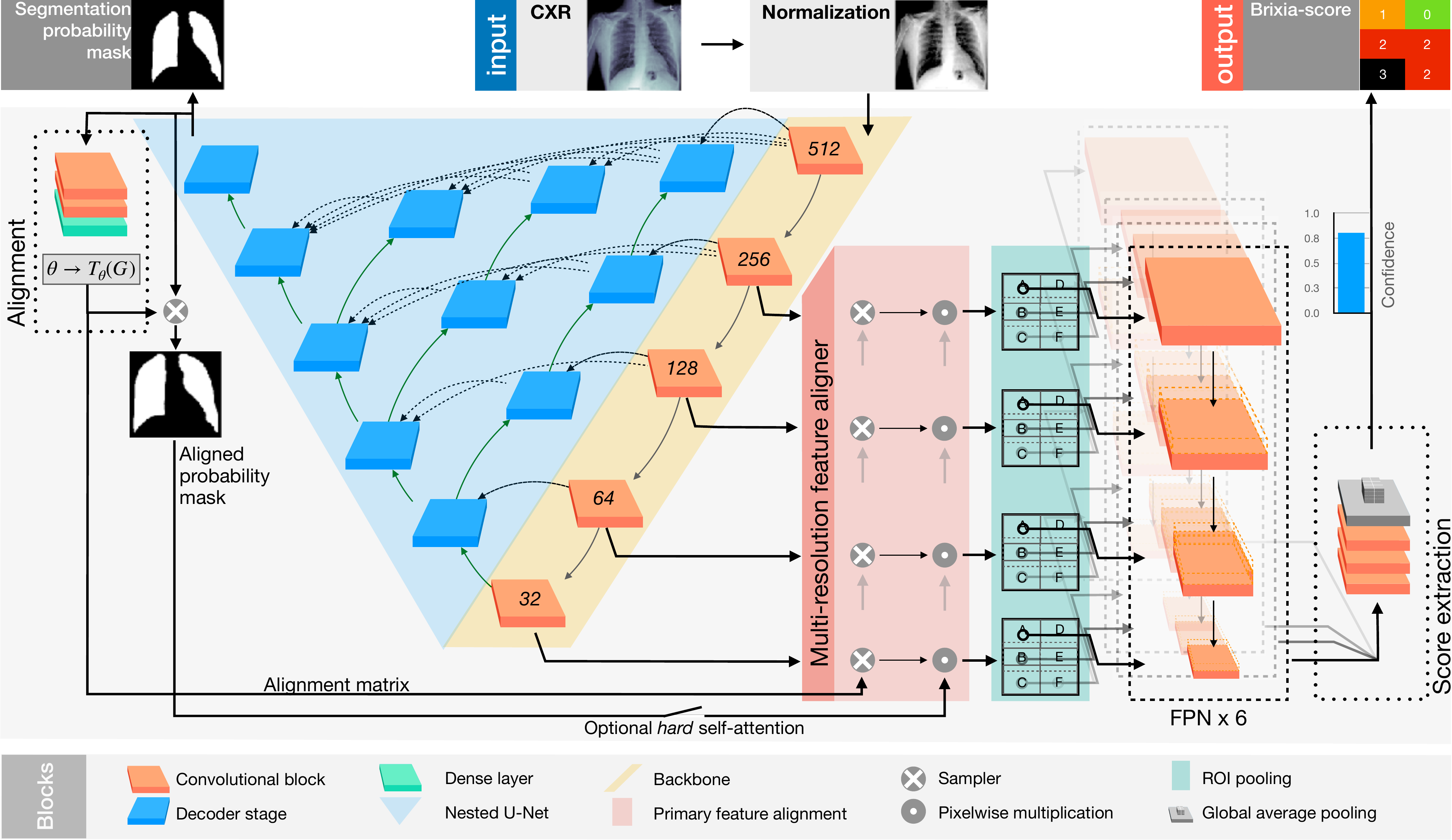}
    \caption{Detailed scheme of the proposed architecture. In particular, in the top-middle the CXR to be analyzed is fed to the network. The produced outputs are: the segmentation mask of the lungs (top-left); the aligned mask (middle-left); the \texttt{Brixia~score} (top-right).}
    \label{fig:global}
\end{figure*}
\begin{table}[h!]
\small
\caption{Brixia COVID-19 dataset details.} 
\centering 
\begin{tabular}{C{3cm}  C{3cm}} 
\toprule
	\textbf{Parameter} & \textbf{Value} \\
\midrule
\midrule
	Modality & CR (62\%) - DX (38\%) \\	\midrule
	View position & AP (87\%) - PA (13\%) \\ \midrule
	Manufacturers & Carestream, Siemens \\	\midrule
	Image size & $ (1056\div3050)\times(1186\div3376)$\\	\midrule
	%No.  of patients & 2400 (???)\\		\midrule
	No.  of  images & 4,703\\	\midrule
	Training set & $3,311$ images\\ 	\midrule
	Validation set & $945$ images\\ 	\midrule
	Test set & $447$ images%\textsuperscript{*} (???)
	\\ \bottomrule \\[-0.25cm]
%\multicolumn{2}{l}{\textsuperscript{*} 50 of which are made publicly available.}
\end{tabular}
\label{tab:dataset} 
\end{table}

\subsubsection{Consensus-based Gold Standard}
\label{BGS}
To assess the level of inter-rater agreement among human annotators, we asked other 4 radiologists to rate a subset of 150 images belonging to the test set of the Brixia COVID-19 dataset.
While $R_0$ is the original clinical annotation in the patient report, we name $R_1, R_2, R_3,$ and $R_4$ the four radiologists that provided additional scores. 
Their expertise is variegated so as to represent the whole staff experience: we have one resident at the 2$^{nd}$ year of training, and three staff radiologists with 9, 15 and 22 years of experience (reported numerical ordering of $R_i$ does not necessarily correspond to seniority order). 
A Gold Standard score is then built based on a majority criterion, by exploiting the availability of multiple ratings (by $R_0, R_1, R_2, R_3,$ and $R_3$), using seniority in case of equally voted scores.
Building such Gold Standard is useful, on the one hand, to grasp the inbuilt level of error in the training set, and on the other hand, to gain a reference measure for human performance and inter-rater variability assessments.

\subsection{Public COVID-19 dataset} 
\label{ssec:public-dataset}

To later demonstrate the robustness and the portability of the proposed solution, we exploit the public repository by \cite{cohen2020covid}, which contains CXR images of patients which are positive or suspected of COVID-19\footnote{We downloaded a copy on May 11th, 2020.}.
This dataset is an aggregation of CXR images collected in several centers worldwide, at various spatial resolutions, and other unknown image quality parameters, such as modality and window-level settings.
In order to contribute to such public dataset, two expert radiologists, a board-certified staff member $R_s$ and a trainee $R_j$, with 22 and 2 years of experience, respectively, produced the related \texttt{Brixia~score} annotations for CXR in this collection, exploiting 
\texttt{labelbox}\footnote{https://labelbox.com/}, an online solution for labelling.
After discarding few problematic cases (e.g., images with a significant portion missing, too low resolution/quality, the impossibility of scoring for external reasons, etc.), the obtained dataset is composed of 192 CXRs, completely annotated according to the \texttt{Brixia~score} system\footnote{Available from both https://github.com/BrixIA/Brixia-score-COVID-19 and https://github.com/ieee8023/covid-chestxray-dataset}. Its Global Score distribution is shown in Figure~\ref{fig:data-distr}~(right).

%% -----------------------------------------
%%     SECTION 5 - METHOD
%% -----------------------------------------

\section{End-to-end multi-network model}
\label{sec:methods}

\subsection{Proposed architecture}
\label{subsec:model}

To predict the pneumonia severity of a given CXR, we propose a novel architecture where different blocks cooperate, in an end-to-end scheme, to segment, align, and predict the \texttt{Brixia~score}. Each block solves one of the specific tasks in which the severity score estimation can be subdivided, while the proposed global end-to-end loss provides the glue that concurs to the creation of a single end-to-end model.
%The result is a single model that estimate a severity score well focused on the lungs.}
The global scheme is depicted in Figure~\ref{fig:global}, while details on single parts follow.

\paragraph{Backbone} The input image is initially processed by a cascade of convolutional blocks, referred to as Backbone (in yellow in Figure~\ref{fig:global}).
This cascade is used both as the encoder section of the segmentation network, and as the feature extractor for the Feature Pyramid Network of the classification branch. 
To identify the best solution at this stage, we tested different backbones among the state-of-the-art, i.e., ResNet \citep{he2016deep}, VGG \citep{simonyan2014very}, DenseNet \citep{huang2017densely}, and Inception \citep{szegedy2017inception}. 

\paragraph{Segmentation}
Lung segmentation is performed by a nested version of U-net, also called U-Net++ \citep{zhou2018}, a specialized architecture oriented to medical image segmentation (in blue in Figure~\ref{fig:global}). 
It is composed of an encoder-decoder structure, where the encoder branch (i.e., the Backbone) exploits a nested interconnection to the decoder. 

\paragraph{Alignment}
The segmentation probability map produced by the Unet++ decoder stage is used to estimate the alignment transformation.
%, which is employed to align both the segmentation mask and the feature pyramid. 
Alignment is achieved through a spatial transformer network \citep{jaderberg2015} able to estimate the spatial transform matrix in order to center, rotate, and correctly zoom the lungs. 
The spatial transformer network estimates a six-valued affine transform that is used to resample (by bilinear interpolation) and align all features at various scales produced by the multi-resolution backbone, before they are forwarded to the ROI (Region Of Interest) Pooling.

\begin{figure}
    \centering
    \includegraphics[width=0.5\textwidth]{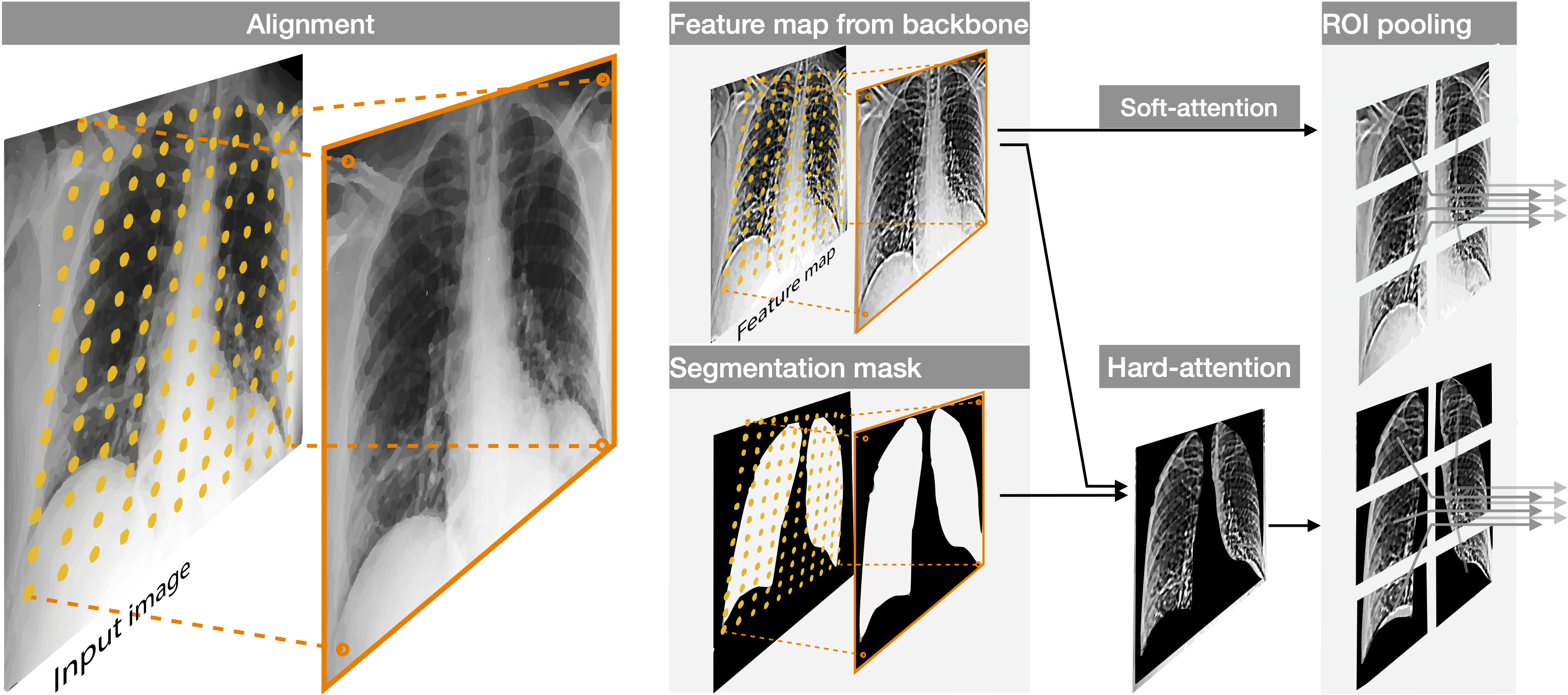}
    \caption{Example of the alignment through the resampling grid produced by the transformation matrix, and its application to both the segmentation mask and the feature maps. On the right, the hard-attention mechanism and the ROI Pooling operation.}
    \label{fig:align-pool}
\end{figure}

This alignment block is pre-trained on the synthetic alignment dataset in a weakly-supervised setting, using a Dice loss. The weakly-supervision is due to the fact that we do not provide transformation matrices as ground-truth labels, but original mask images before the synthetic transformation. Moreover, the labels are anyway noisy since images in the segmentation dataset (that compose the base for the used synthetic database) may be not perfectly aligned. More precisely we assume that the level of misalignment already present in dataset images is in general negligible with respect to the transforms we artificially apply. This allows a meaningful pre-training of the alignment block, as will be shown in the results, able to compensate the inaccurate patient positioning often present due to the various critical conditions of CXR acquisitions.
%in the context of the COVID-19 pandemic.}
%produces a self-attentive mechanism useful to propagate the correct area of the lungs, through the ROI pooling, towards the final classification stage. \textcolor{orange}{We named this block "multi-resolution feature aligner", and can be seen in Figure \ref{fig:global}.}
In Figure \ref{fig:global} we show the action of the alignment on the features at various scales generated by the backbone in a block we call ``multi-resolution feature aligner", while in Figure~\ref{fig:align-pool} we give a representation of this block, where aligned features are produced by bi-linear resampling on the original maps. This resampling scheme is not only used to align the backbone features, but can be also helpful for realigning the segmentation map in an optional hard-attention configuration, as explained in the following.

\paragraph{Hard vs. Soft self-attention}
A hard self-attention mechanism can be applied by masking the aligned features with the segmentation mask (obtained as softmax probability map). 
This option has the advantage of switching off possible misleading regions outside the lungs, favouring the flow of relevant information only. 
Therefore the network can operate in two configurations: either with a hard self-attention scheme (HA), in which the realigned soft-max segmentation mask (from 0 to 1) is used as a (product) weighting mask in the multi-resolution feature aligner, or with a soft version (SA), where the segmentation mask is only used to estimate the alignment transform, but not to mask the aligned backbone features. 

\paragraph{ROI Pooling}
The aligned (and optionally hard masked) features are finally used to estimate the $3\times2$ matrix containing the \texttt{Brixia~score}.
To this purpose, ROI Pooling is performed on a fixed grid with the same dimensions. 
In particular, from the aligned features map produced by the multi-resolution feature aligner, the ROI Pooling extracts the 6 \texttt{Brixia score} regions (with a vertical overlap of 25\%, and no horizontal overlap, since the left/right boundary between lungs is easily identified, while the vertical separation presents a larger variability).
This pooling module introduces \textit{a-priori} information regarding the location of the six regions (i.e, $A, B, \ldots, F$), while leaving to the network the role to correctly rearrange the lungs feature maps by means of the alignment block. As ouput, this block returns 6 feature maps (one for each lung region) for each level in the backbone. 
The combination of alignment and ROI pooling produces (especially with hard attention) a self-attentive mechanism useful to propagate only the correct area of the lungs towards the final classification stage.

%\hl{@MS: ogni volta che mi hai spiegato questo blocco e il successivo, spiegandomi Fig.3, le tue spiegazioni mi sembravano molto piu' ricche di quanto scritto qui, che evidentemente non basta a comprendere le figure. Riesci ad arricchire (per evidenziare meglio i tuoi contributi) e nel contempo migliorare la comprensibilita'?} DONE

\paragraph{Scoring head}
The final scoring module exploits the idea of Feature Pyramid Networks (FPN) \citep{lin2017cvpr}  for the combination of multi-scale feature maps.
As depicted in Figure~\ref{fig:global}, we combine feature maps that come from various levels of the network, therefore with different semantic information at various resolutions. 
The multi-resolution feature aligner produces input feature maps that are well focused on the specific area of interest.
Eventually, the output of the FPN layer flows in a series of convolutional blocks to retrieve the output map ($3\times 2\times 4$, i.e., 3 rows, 2 columns, and 4 possible scores $[0,\ldots, 3]$). 
The classification is performed by a final Global Average Pooling layer and a SoftMax activation.

\subsection{Loss Function and Model Training}
\label{ssec:training}

The Loss function we use for training, is a sparse categorical cross entropy ($\mathrm{SCCE}$) with a (differentiable) mean absolute error ($\mathrm{MAE}^d$) contribution:

\begin{equation}
\label{eq-loss}
    \textsl{L} = \alpha \cdot \mathrm{SCCE} + (1 - \alpha) \cdot \mathrm{MAE}^d
\end{equation}
where $\alpha$ controls how much weight is given to $\mathrm{SCCE}$ and $\mathrm{MAE}^d$, which are defined as follows:

\begin{equation}
    \mathrm{SCCE} = - \frac{1}{C} \sum_{c} y_{c} \log (\hat{y}_{c})
\end{equation}
\begin{equation}
    \mathrm{MAE}^d = \frac{1}{C} \left \| y - \sum_{c\in C} \frac{e^{\beta \hat{y}_{c}}}{\sum_{j \in C} e^{\beta \hat{y}_{j}}}c  \right \|
\end{equation}

where $y$ is the reference \texttt{Brixia~score}, $\hat{y}$ is the predicted one, and $c\in[0,1,\dots3]$ is the score class.
To make the mean absolute error differentiable ($\mathrm{MAE}^d$), $\beta$ can be chosen to be an arbitrary large value.

The selection of such loss function is coherent with the choice to configure the \texttt{Brixia~score} problem as a joint multi-class classification and regression.
Tackling our score estimation as a classification problem allows to associate to each score a confidence value: this can be useful to either produce a weighted average, or to introduce a quality parameter that the system, or the radiologist, can take into account.
Moreover, the $\mathrm{MAE}^d$ component, while being meaningful for the scoring system, adds robustness to outliers and noise.

Due to the nature and complexity of the proposed architecture, the training of network weights takes place at several stages, according to a \textit{from-the-part-to-the-whole} strategy, according to a task-driven policy. This does not contradict the end-to-end nature of the system. Indeed, not only pre-training each sub-network in a structured multi-level training is a possibility, but it is also the most advisable way to proceed, as also recognized in \citep{glas2017}.
The different sections of the overall network are therefore first pre-trained on specific tasks: U-Net++ is trained using the lung segmentation masks in the segmentation datasets; the alignment block is trained using the synthetic alignment dataset (in a weakly-supervised setting); the classification portion (Scoring head) is trained using Brixia COVID-19 dataset, while blocking the weights of the other components (i.e., Backbone, Segmentation, Alignment).
Then, a complete fine-tuning on Brixia COVID-19 dataset follows, making all weights (about 20 Million) trainable.
The network hyper-parameters eventually undergo a selection that maximizes the MAE score on the validation set. 

\subsection{Implementation details}
The network has an input size of $512 \times 512$. 
The selected backbone is a ResNet-18 \cite{he2016deep}, because it offers the best trade-off between the expressiveness of the extracted features and the memory footprint (as in the case of the input size). 
In the network, we use the rectified linear unit (ReLU) activation functions for the convolutional layer of the backbone, and the U-Net++, while the Swish activation function by \cite{ramachandran2017searching} is used for the remaining blocks. 
We extensively make use of online augmentation throughout the learning phases. 
In particular, we apply all the geometric transformations described in Section \ref{ssec:data-align}, plus random brightness and contrast, as well. 
Moreover, we randomly flip images horizontally (and the score, accordingly) with a probability of 0.5.
We exploit, for training purposes, two machines equipped with Titan\textsuperscript{\textregistered} V GPUs.
We train the model by jointly optimizing the sparse categorical cross-entropy function and MAE, with a selected $\alpha=0.7$. 
Convergence is achieved after roughly $6$hours of training ($80$ epochs), using Adam \citep{adam2014} with an initial learning rate of $3\cdot 10^{-2}$, halving it on flattening.
The batch size is set to $8$.
%Other technical details not included here for the sake of readability, can be found either in Figure~\ref{fig:global}, or in the documentation accompanying the source code distribution.

\subsection{Super-pixel explainability maps}
\label{sec:explain}
To evaluate whether the network is predicting the severity score on the basis of correctly identified lung areas, we need a method capable of generating explainability maps with a sufficiently high resolution.
Unfortunately, with the chosen network architecture, the popular Grad-CAM approach \citep{selvaraju2017}, and similar ones, generates poorly localized, and spatially blurred, visual explanations of activation regions, as it happens for example also in \citep{karim2020deepcovidexplainer,oh2020deep}.
Moreover, we would face a concrete difficulty in our context as Grad-CAM would generate 6(regions) $\times$ 4(classes) maps that would be difficult to combine for an easy and fast inspection by the radiologist during the diagnosis.
%, where a patch-based solution (although far from our approach) resulted beneficially.
For these reasons, we designed a novel method for generating useful explainability maps, loosely inspired by the LIME \citep{ribeiro2016} initial phases.
The creation of the explainability map starts with the input image division into $N$ super-pixels i.e., regions that share similar intensity and pattern, extracted as in \cite{vedaldi2008quick}. 
Starting from the input image, we create $N$ image replicas in which a single super-pixel $i$ (from $1$ to $N$) is masked to zero. 
We call $p_0$ the probability map that the model produces starting from the original image ($2\times 3\times 4$ values, \textit{i.e.}, one for each of the 4 severity classes in every lung sector that composes the \texttt{Brixia~score}). We call instead $p_i$ the probability map produced from the $i^{th}$ replica.
%For each of the $N$ generated super-pixel, we create an image with the corresponding super-pixel set to zero value (super-pixel-masked image), and for such $N$ images we predict the six-valued \texttt{Brixia~score} (thus excluding the specific super-pixel from the computation).
%We then accumulate the differences with respect to the predicted values considering the whole original images for each score class on the specific super-pixel.
We then accumulate for the differences between all the super-pixel masked predictions $p_i$ and the original prediction $p_0$.
Given $S$ the set of super-pixels, the output explanation map $E$ is obtained as:
\begin{equation}
    E = \sum_i^{|S|} S_i \cdot  ( p_i - p_0 )
\end{equation}
%where $p_0$ is the probability map obtained with the original input, while $p_i$ is the probability associated to the super-pixel-masked image.
Intuitively, the obtained maps highlight the regions that most account for the score in exam.
Examples can be appreciated in Figure~\ref{fig:framework} (bottom-right) as well as in Figure~\ref{fig:results}: the more intense the color, the more important the region contribution to the score decision.

%% -----------------------------------------
%%          SECTION 6 - RESULTS
%% -----------------------------------------

\section{Results}
\label{sec:results}

Through an articulated experimental validation we first consider how single components of the architecture operate (Sections~\ref{res:seg} and \ref{res:alig}) and we give a complete picture of the severity assessment performance referred to the whole collected dataset (Section~\ref{res:score}). Then we deal with the inter-rater variability issue and demonstrate that the proposed solution overcomes the radiologists' performance on a consensus based Gold Standard reference (Section~\ref{res:gold}). We then widen the scope of our results in two directions: 1) we consider other proposed severity scores and give measures, observations, and comparisons that clearly support the need of dedicated solutions to the COVID-19 severity assessment (Section~\ref{res:other-scores}); 2) we consider the portability (both direct or mediated by a fine-tuning) of our model on public data collected in the most different locations and conditions, verifying a high degree of robustness and generalization of our solution (Section~\ref{res:portab}). Qualitative evidences of the role that the new explainability solution can have in a responsible and transparent use of the technology are then proposed (Section
~\ref{res:explain}). We conclude with some ablation studies and technology variation experiments aiming essentially at showing how the complexity of our multi-network architecture is neither over- nor under-sized, but it is adequate to the needs and complexity of the target visual task (Section~\ref{res:ablation}).
All results presented in this section are discussed in the following Section~\ref{sec:disc}. 

\subsection{Lung segmentation}
\label{res:seg}

The performance of the segmentation stage shows totally comparable results with respect to state-of-the-art methods \citep{candemir2019}, both DL based \citep{zhou2018,frid2018improving,oh2020deep}, and hybrid \citep{candemir2014}. 
Table \ref{tab:segmentation} reports the results for the U-Net++ (and U-Net comparison) in terms of Dice coefficient and Intersection over Union (IoU), aka Jaccard index, obtained on the test set of the segmentation datasets (Sec.\ref{ssec:data-seg}).
Training curves for the segmentation task (on both training set, and validation set), tracking the IoU, are shown in Figure~\ref{fig:tc}-a for BS-Net in HA configuration.

\begin{table}[ht]
\small
\centering
\caption{Lung segmentation performance.}
\begin{tabular}{@{}llll@{}}
\toprule
       & Backbone & Dice coefficient & IoU   \\ \midrule
U-Net++ & ResNet-18 & 0.971            & 0.945 \\
U-Net   & ResNet-18 & 0.969            & 0.941 \\ \bottomrule
\end{tabular}
\label{tab:segmentation}
\end{table}

\begin{table*}[ht]
\small
\centering
\caption{\texttt{Brixia~score} prediction performance parameters for the four considered models on the Brixia COVID-19 dataset (only blind test set results reported). Parameters are evaluated on each single lung region (A-F), averaged on all the lung regions and on the Global Score (P-value $\ll 0.00001$ everywhere).}
\label{tab:disparity}
\begin{tabular}{@{}llllllll|l|l@{}}
          &                             & A        & B        & C        & D        & E        & F        & \shortstack[l]{Avg. on\\ regions} & \shortstack[l]{Global\\ score} \\ \midrule
BS-Net-Ens  & \multirow{4}{*}{MEr}      & 0.169 & -0.038 & -0.056 & 0.125 & -0.045 & -0.192 & \textbf{-0.006}          & \textbf{-0.036}  \\
BS-Net-SA &                             & 0.107 & -0.087 & -0.171 & 0.082 & -0.129 & -0.343 & -0.090          & -0.541  \\
BS-Net-HA &                             & 0.156 & -0.147 & -0.085 & 0.107 & -0.016 & -0.238 & -0.037          & -0.223  \\
ResNet-18  &                             & 0.356 & -0.038 & -0.056 & 0.125 & -0.045 & -0.192 & 0.100           & 0.601   \\ \midrule
BS-Net-Ens  & \multirow{4}{*}{MAE} & 0.459 & 0.448  & 0.412  & 0.374 & 0.459  & 0.494  & \textbf{0.441}           & \textbf{1.728}   \\
BS-Net-SA &                             & 0.499 & 0.501  & 0.506  & 0.408 & 0.499  & 0.566  & 0.496           & 1.846   \\
BS-Net-HA &                             & 0.481 & 0.477  & 0.481  & 0.370 & 0.488  & 0.532  & 0.471           & 1.826   \\
ResNet-18  &                             & 0.543 & 0.486  & 0.506  & 0.452 & 0.584  & 0.530  & 0.517           & 1.951   \\  \midrule
BS-Net-Ens  & \multirow{4}{*}{SD}        & 0.604 & 0.524  & 0.540  & 0.541 & 0.574  & 0.609  & \textbf{0.565}           & \textbf{1.429}   \\
BS-Net-SA &                             & 0.638 & 0.560  & 0.579  & 0.576 & 0.594  & 0.634  & 0.597           & 1.514   \\
BS-Net-HA &                             & 0.613 & 0.575  & 0.583  & 0.552 & 0.594  & 0.616  & 0.589           & 1.505   \\
ResNet-18  &                             & 0.657 & 0.579  & 0.591  & 0.632 & 0.657  & 0.601  & 0.619           & 1.710   \\  \midrule
BS-Net-Ens  & \multirow{4}{*}{CC}       & 0.675 & 0.779  & 0.731  & 0.682 & 0.737  & 0.672  & \textbf{0.713}           & \textbf{0.862}   \\
BS-Net-SA &                             & 0.635 & 0.733  & 0.675  & 0.633 & 0.718  & 0.636  & 0.672           & 0.847   \\
BS-Net-HA &                             & 0.665 & 0.742  & 0.679  & 0.662 & 0.722  & 0.645  & 0.686           & 0.845   \\
ResNet-18  &                             & 0.598 & 0.739  & 0.667  & 0.562 & 0.655  & 0.643  & 0.644           & 0.842  \\
\bottomrule
\end{tabular}
\end{table*}

\subsection{Alignment}
\label{res:alig}
After training on the synthetic dataset described in Section~\ref{ssec:data-align}, we report the following alignment results: Dice coefficient = 0.873, IoU = 0.778. The Dice coefficient and IoU are calculated using the classical definition by considering the original masks (before synthetic misalignment) and the ones re-aligned by means of the affine transform estimated by the network.
Training curves for the alignment task (on both training set, and validation set), tracking the IoU, are shown in Figure~\ref{fig:tc}-b, always for HA configuration.
Convergence behaviors are clearly visible, with a consolidating residual distance between training and validation curves. 

Despite the difficulty of the task and the fact that simulated transforms are usually overemphasised with respect to misalignment found in real data, %this apparently fair performance is fully satisfactory in our context.
the measured alignment performance on the synthetic dataset produces a significant performance boost, as it clearly emerges from the ad-hoc ablation studies on rotations (see test in Section \ref{res:ablation}).

Residual errors are typically in the form of slight rotations and zooms, which are not critical and well tolerable in terms of overall self-attentive behavior. 
Moreover, during a visual check of the Brixia Covid-19 test dataset, no impairing misalignments (lung outside the normalized region) were observed.

\subsection{Score prediction on Brixia COVID-19 dataset}
\label{res:score}

To evaluate the overall performance of the network, we analyze the \texttt{Brixia~score} predictions with respect to the score assigned by the radiologist(s) $R_0$, i.e. the one(s) who originally annotated the CXR during the clinical practice.
Discrepancies found on the 449 test images of Brixia COVID-19 dataset are evaluated in terms of Mean Error (MEr), Mean Absolute Error (MAE) with its standard deviation (SD), and Correlation Coefficient (CC). 

Four networks are considered for comparison, three of which are different configurations of BS-Net: the hard attention (HA) one, the soft attention (SA) one, and one ensemble of the two previous configurations (ENS).
In particular ENS configuration exploits both HA and SA paths to make the final prediction (by averaging their output probabilities): these realizations are the best model with respect to the validation set, and the obtained model after the last training iteration.
The fourth compared network is ResNet-18 (the backbone of our framework) as an all-in-one solution (with no dedicated segmentation, nor alignment stage), since it is one among the most adopted architectures in studies involving CXR analyses, also until today in the COVID-19 context \citep{minae2020}.

\begin{figure*}
    \centering
    \includegraphics[width=1.00\textwidth]{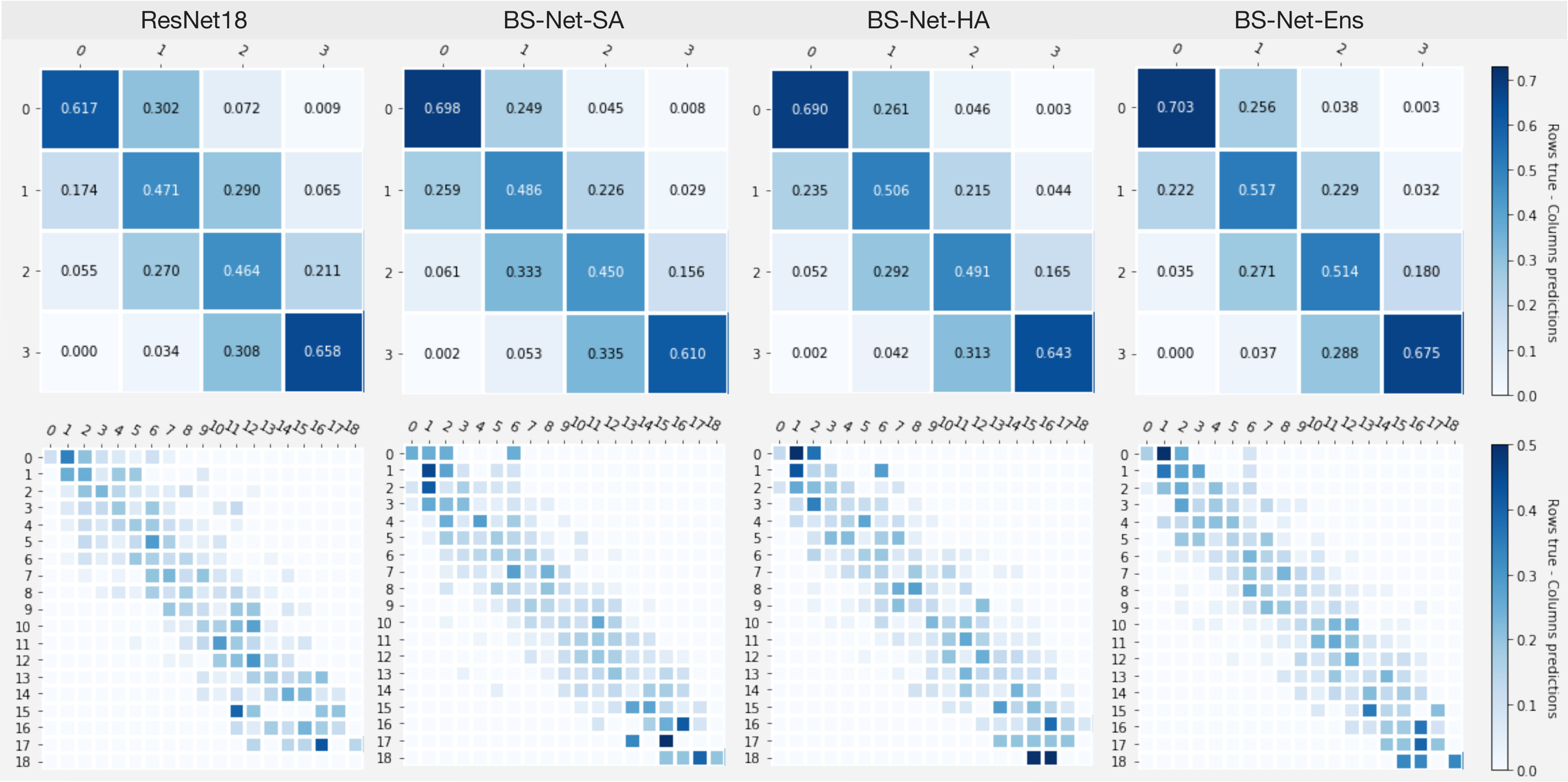}
    \caption{Consistency/confusion matrices based on lung regions score values (top, 0--3), and on Global Score values (bottom, 0--18).}
    \label{fig:cms}
\end{figure*}

Table~\ref{tab:disparity} lists all performance values referred to each of the six regions (A to F) of the \texttt{Brixia~score} (range [0-3]), to the average on single regions, and to the Global Score (range [0-18]). 
In a consistency assessment perspective, Figures \ref{fig:cms} (top) and (bottom) show the confusion matrices for the four networks related to the score value assignments for single lung regions, and for their sum (Global Score), respectively.
From Table~\ref{tab:disparity} it clearly emerges that the ensemble decision strategy (ENS) succeeds in combining the strengths of the soft and hard attention policies, with the best average MAE on the six regions of 0.441.
Conversely, the straightforward end-to-end approach offered by means of an all-in-one ResNet-18 is always the worst option compared to the three BS-Net configs.

\begin{figure}
    \centering
    \includegraphics[width=0.4\textwidth]{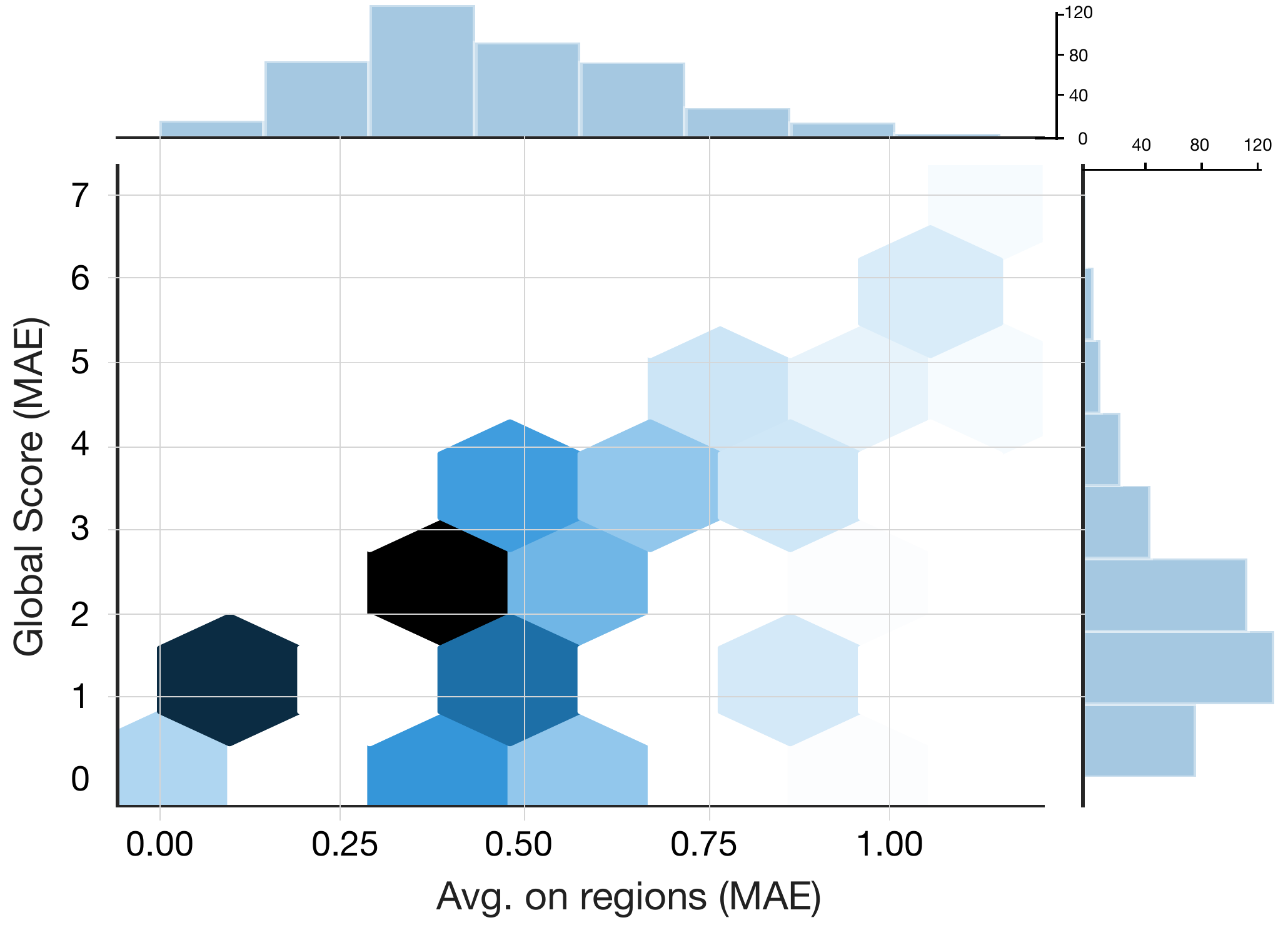}
    \caption{Single and joint MAE distribution for lung regions and Global Score predictions obtained by BS-Net (ENS).}
    \label{fig:err-distr2}
\end{figure}

\begin{figure*}[t]
    \centering
    \includegraphics[width=0.98\textwidth]{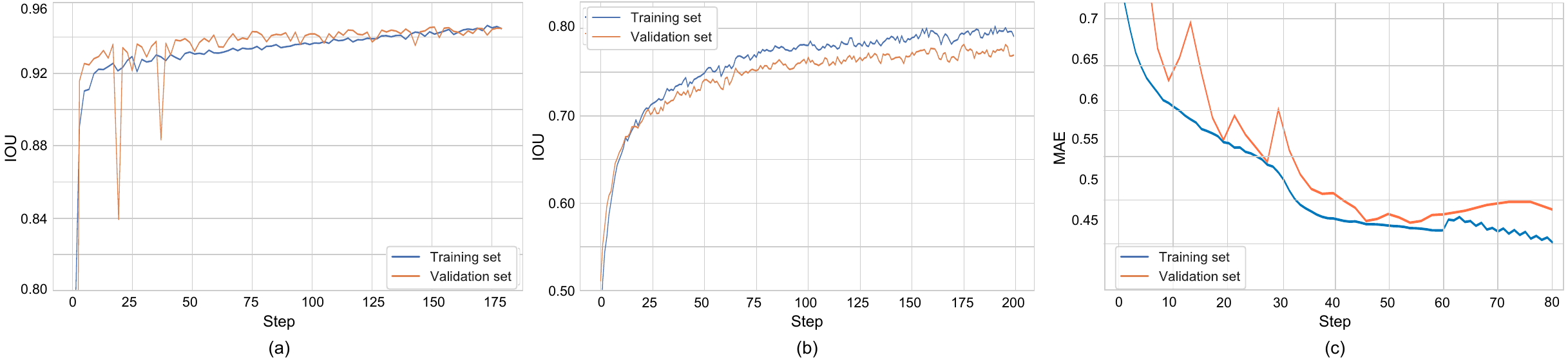}
    \caption{Training curves related to BS-Net-HA. Segmentation (a);  Alignment (b); \texttt{Brixia~score} prediction -- best single model (c).}
    \label{fig:tc}
\end{figure*}

The error distribution analysis on MAE, depicted in Figure~\ref{fig:err-distr2}, shows the prevalence of lower error values on both single lung regions and Global Score estimations. 
The joint view gives another evidence that single-region errors unlikely sum up as constructive interference. 
Training curves for this prediction task (on both training set, and validation set), tracking the Mean Absolute Error (MAE) on the set of lung sectors are shown in Figure~\ref{fig:tc}-c (HA configuration).
Convergence behaviors are clearly visible, with a consolidating residual distance between training and validation curves.

\subsection{Performance assessment on Consensus-based Gold Standard}
\label{res:gold}
Results in Table~\ref{tab:disparity} are computed with respect to the score assigned by a single radiologist (i.e., $R_0$) among the ones in the whole staff.
With the aim of providing a more reliable reference, we consider the Consensus-based Gold Standard (CbGS) dataset (Section \ref{BGS}). This allows to recompute the BS-Net (ENS) performance on a subset of 150 images, which are annotated with multiple ratings (by $R_0, R_1, R_2, R_3$, and $R_4$) from which a consensus reference score is derived.
Table~\ref{tab:interrater} (top-center) clearly shows a significant improvement on MAE with respect to the comparison versus $R_0$ only (0.424 vs. 0.452).
%This further demonstrates the quality and robustness of the \texttt{Brixia~score} estimation, despite the inter-rater variability issues.

\begin{table}[t]
\centering
\caption{Results on the Consensus-based Gold Standard dataset (150 images):\\ (top) performance of BS-Net (ENS) computed on the Consensus-based Gold Standard; (center) performance of BS-Net (HS) vs original rater $R_0$; (bottom) averaged performance of all $R_i$ vs CbGS}
\label{tab:interrater}
\small
\begin{tabular}{@{}lllll@{}}
\toprule
                & MEr    & MAE   & SD    & CC   \\ \midrule
& \multicolumn{4}{c}{BS-Net (ENS) vs. CbGS} \\ \cmidrule{2-5}
Avg. on reg. & -0.133 & \textbf{0.424} & 0.580 & 0.743 \\
Global score         & 0.800  & 1.787 & 1.354 & 0.907 \\
\midrule 

& \multicolumn{4}{c}{BS-Net (ENS) vs. $R_0$} \\ \cmidrule{2-5}
Avg. on reg. & -0.019 & 0.452 & 0.575 & 0.754 \\
Global score         & -0.113   &  1.847 & 1.553  & 0.834 \\ 

\midrule
& \multicolumn{4}{c}{Average on all pairs of radiologists} \\ \cmidrule{2-5}
Avg. on reg. & -0.131 & 0.528 & 0.614 & 0.736 \\
Global score         & -0.784  &  2.592 & 1.965  & 0.835 \\ 
\bottomrule
\end{tabular}
\end{table}

\begin{figure}[t]
    \centering
    \includegraphics[width=0.48\textwidth]{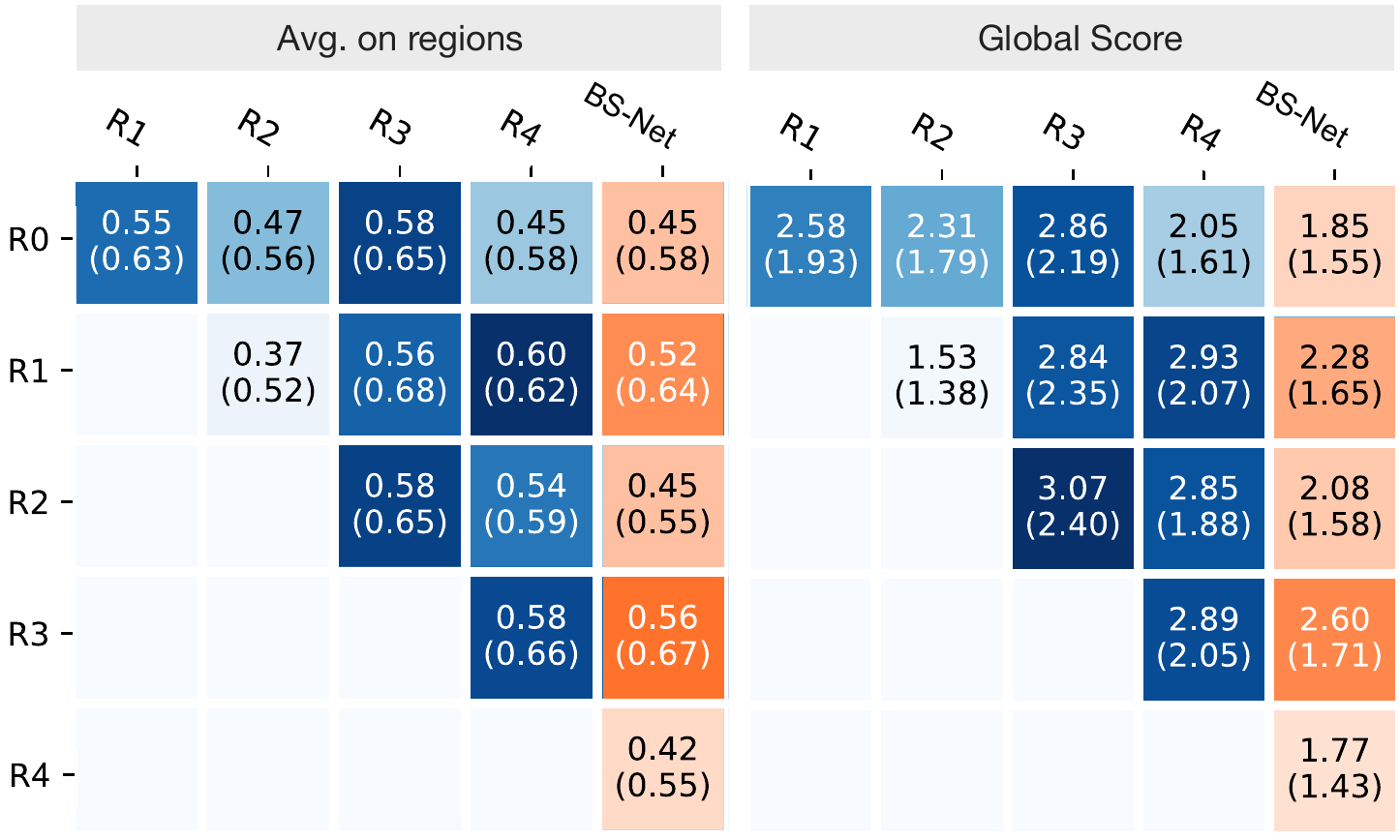}
    \caption{Pairwise inter-rater results in terms of MAE (and SD). In the most right column (orange), the inter-rater results with predictions by BS-Net-Ens.}
    \label{fig:inter-rater}
\end{figure}

\paragraph{Inter-rater agreement: human vs. machine performance}
In Figure~\ref{fig:inter-rater} we assess the inter-rater agreement by listing MAE and SD values referred to all possible pairs of raters, including $R_0$, and also BS-Net (ENS), as a further ``virtual rater''. 
Looking at how BS-Net behaves (orange boxes in Figure~\ref{fig:inter-rater}), we observe the level of agreement reached between the network and any other rater is, almost always, higher than the inter-rater agreements between any pair of human raters.
For example, by considering $R_0$ as a common reference, we have an equal performance only in the case of $R_4$.
Table~\ref{tab:interrater} confirms how BS-Net (top) performs on average significantly better (MAE 0.424) with respect to the global indicators (bottom) coming from averaging all pairwise $R_i$ vs $R_j$ ($i\neq j$) comparisons (MAE 0.528).

The inter-rater agreement values (between human raters) indicates a moderate level of agreement (both averaged two-raters Cohen's kappa value and multi-rater Fleiss' kappa value, based on single cell scores, are around 0.4). 
%\hl{spiegare che qs performance vanno interpretate buone visto che abbiamo una weak supervision (cosa che  dovrebbe essere stata chiarita altrove), ma che si dimostraranno adeguate quanto andremo a testare il contributo della parte di allineamento sulle performance generali @Alby, io non ricordo di aver scritto questo commento. E' tuo? Sergio}
One the one hand, this is a confirmation that the \texttt{Brixia~score}, probably succeeds in reaching the tough compromise between maximizing the score expressiveness (spatial and rating granularity), while keeping under control the level of  subjectivity (inter-rater variability). On the other hand, measured inter-rater variability levels constitute a clear limitation that bound the network learning abilities (weak supervision), concurrently allowing to assess whether the network surpasses single radiologists performance. 
The opportunity for the network to learn not only from a single radiologist but virtually from a fairly large community of specialists (being $ R_0 $ a varying radiologist) is, de facto, the margin that we have been able to exploit.

%The fact that inter-rater agreement values between human raters are indicative of a fair to moderate level of agreement (averaged two-raters Cohen's kappa value and the multi-rater Fleiss' kappa value, based on single cell scores, around 0.4) clearly constitute the limitation that bound the network learning abilities.

%\subsection{Performance assessment on other severity scores}
\subsection{Performance assessment on Toussie score}
\label{res:other-scores}
In Table \ref{tab:toussie} we show the performance related to six-valued \texttt{T~score}, akin to the score presented in \cite{toussie2020}, derived by thresholding the \texttt{Brixia~score} as described in Section \ref{ssec:OtherSeverity}.
We provide results on both the whole Brixia COVID-19 test set and on the Consensus-Based Gold Standard set. Interestingly, the correlation increases from 0.73 on the whole test set to 0.81 on the CbGS set, while MAE, despite maintaining acceptable values, increases as well (probably due to a residual from a non perfect mapping between the two scores).

\subsection{Performance assessment on GE-LO score}
In Table \ref{tab:ch-opacity} we simulate the computation of the \texttt{LO~score} by \texttt{Brixia~score} mapping followed by linear regression (see Sec.\ref{ssec:OtherSeverity}). Thanks to annotations we provided for CXRs in the dataset by \citep{cohen2020data} we had the possibility to perform a direct comparison with the non-specific method described in \cite{cohen2020predicting} from which we report only the best results. We report the results produced by BS-Net-ENS on the intersection between the subset of the Cohen dataset considered in our work (Section \ref{ssec:public-dataset}) and the CXRs we found used in \cite{cohen2020predicting} (a retrospective cohort of 94 PA CXR images). We also produce the results considering the whole subset for which we produced \texttt{Brixia~score} annotations (obtaining virtually equivalent results). We also report \texttt{LO~score} related results starting from \texttt{Brixia~score} assigned by our experts. The performance boost produced by a prediction from a specifically designed solution is evident, and this is coherent also to the considerations and limitations acknowledged by the authors of \cite{cohen2020predicting}.

\begin{table}
\centering
\caption{\texttt{T~score} performance\\ (faithful simulation of the score proposed in \citep{toussie2020}).}
\label{tab:toussie}
\small
\begin{tabular}{@{}lllll@{}}
\toprule
             & MEr           & MAE          & SD          & CC         \\ \midrule
             & \multicolumn{4}{l}{BS-Net-ENS on the whole test set}        \\ \cmidrule(l){2-5} 
Avg. on reg. & 0,009         & 0,147        & 0,349       & 0,51       \\
Global score & 0,056         & 0,742        & 0,921       & 0,728      \\ \midrule
             & \multicolumn{4}{l}{BS-Net-ENS on the CbGS set} \\ \cmidrule(l){2-5} 
Avg. on reg. & -0,074        & 0,174        & 0,38        & 0,563      \\
Global score & -0,447        & 0,94         & 1,07        & 0,81       \\ \bottomrule
\end{tabular}
\end{table}

\begin{table*}
\centering
\caption{Opacity score linear regression from global \texttt{Brixia~score}.}
\label{tab:ch-opacity}
\begin{tabular}{lllllll}
\toprule
{} &    Correlation &          $R^2$ &            MAE &            MSE &          coef &     intercept \\
{} \\
from              &                &                &                &                &                &                \\
\midrule
Best \citep{cohen2020predicting}  &  0.80$\pm$0.05 &  0.60$\pm$0.17 &  1.14$\pm$0.11 &  2.06$\pm$0.34 &  -- &  -- \\
BS-Net on Cohen set  &  0.84$\pm$0.05 &  0.54$\pm$0.17 &  0.67$\pm$0.10 &  0.68$\pm$0.18 &  0.31$\pm$0.01 &  0.15$\pm$0.08 \\
BS-Net on our subset &  0.85$\pm$0.08 &  0.53$\pm$0.24 &  0.67$\pm$0.12 &  0.67$\pm$0.22 &  0.31$\pm$0.01 &  0.09$\pm$0.09 \\
$R_s$ on our subset     &  0.90$\pm$0.06 &  0.72$\pm$0.13 &  0.55$\pm$0.10 &  0.45$\pm$0.15 &  0.28$\pm$0.01 &  0.55$\pm$0.07 \\
\bottomrule
\end{tabular}
\end{table*}

% \begin{table}
% \centering
% \caption{Geographic score linear regression from global \texttt{Brixia~score}}
% \label{tab:ch-geographic}
% \begin{tabular}{lllllll}
% \toprule
% {} &    Correlation &          $R^2$&            MAE &            MSE &          coef &      intercept \\
% {} \\
% from              &                &                &                &                &                &                 \\
% \midrule
% Best \citep{cohen2020predicting}  &  0.78$\pm$0.04 &  0.58$\pm$0.09 &  0.78$\pm$0.05 &  0.86$\pm$0.11 &  -- &  -- \\
% BS-Net Cohen set  &  0.87$\pm$0.05 &  0.65$\pm$0.14 &  0.89$\pm$0.14 &  1.34$\pm$0.39 &  0.50$\pm$0.01 &  -0.41$\pm$0.10 \\
% BS-Net our subset &  0.86$\pm$0.08 &  0.58$\pm$0.25 &  0.92$\pm$0.21 &  1.56$\pm$0.69 &  0.50$\pm$0.02 &  -0.43$\pm$0.12 \\
% GT our subset     &  0.92$\pm$0.05 &  0.77$\pm$0.13 &  0.77$\pm$0.14 &  0.95$\pm$0.37 &  0.45$\pm$0.01 &   0.31$\pm$0.08 \\
% \bottomrule
% \end{tabular}
% \end{table}

\subsection{Public COVID-19 datasets: portability tests}
\label{res:portab}

The aggregate public CXR dataset \citep{cohen2020covid}, described in Section~\ref{ssec:public-dataset}, has been judged as inherently well representative with respect to the various manifestations degrees of COVID-19.
This dataset is quite heterogeneous and of a different nature with respect to our dataset acquired in clinical conditions (see Section \ref{ssec:public-dataset}). In particular the non-Dicom format, and the presence of low-resolution images and screenshots, make this dataset a challenging test bed to assess model portability and simulate a worst-case (or stress test) scenario.
%For this reason, it represents a good test bed to assess model portability. 
Exploiting the two independent \texttt{Brixia~score} annotations of this dataset (one from a senior $R_s$, and another from a junior radiologist $R_j$), we performed a portability study, with the aim of deriving some useful guidance for extended use of our model on data generated in other facilities.
In particular, we carried out three tests on BS-Net (HA configuration) measuring the performance on: 1) the whole set of 192 annotated images (full); 
2) a reduced test set with fine-tuning, after random 75/25 splitting (subset, fine-tuning);
2) a reduced test set with partial retraining of the network, after random 75/25 splitting, with segmentation and alignment blocks trained on their specific datasets (subset, from scratch).
In reporting results, we considered the senior radiologist $R_s$ as the reference in order to have the possibility to assess the second rater $R_j$ performance the same way we assess the network performance.
Table~\ref{tab:public-test} lists all results from the above described tests.

\begin{table*}[t]
\centering
\small
\caption{Portability tests on the public dataset \citep{cohen2020covid}. MAE and its SD are listed for both reporting radiologist $R_j$ and BS-Net-HA. The network has been used in three training conditions: 1) as is, originally trained on the Brixia COVID-19 dataset (full), 2) fine-tuned on the public dataset (subset, fine-tuning), and 3) completely retrained, classification part only, on the public dataset (subset, from scratch).}
\begin{tabular}{@{}lllllllll|l|l@{}}
                            &                      & Test   & A     & B      & C      & D     & E      & F      & \shortstack[l]{Avg. on\\ regions}  & \shortstack[l]{Global\\ score}  \\ \toprule
\multirow{2}{*}{Radiologist $R_j$}  & \multirow{6}{*}{MEr} & full    & 0.135 & 0.182  & 0.156  & 0.115 & 0.177  & 0.021  & 0.131           & 0.786        \\
                            &                      & subset & 0.149 &	0.234 &	0.191 &	0.000 &	0.191 &	0.021   &	0.131         &	0.787           \\
\multirow{2}{*}{BS-Net-HA} &                       & full    & 0.177 & -0.167 & -0.177 & 0.167 & -0.208 & -0.651 & -0.143          & -0.859       \\
                            &                      & subset & 0.125 & -0.104 & -0.188 & 0.167 & -0.063 & -0.583 & -0.108          & -0.646        \\
* from scratch              &                      & subset & 0.208 & 0.396  & 0.104  & 0.250 & 0.042  & 0.063  & 0.177           & 1.063         \\
* fine-tuning               &                      & subset & 0.146 & 0.167  & -0.042 & 0.229 & 0.167  & -0.208 & 0.076           & 0.458         \\ \midrule
\multirow{2}{*}{Radiologist}  & \multirow{6}{*}{MAE} & full    & 0.396 & 0.401  & 0.438  & 0.333 & 0.396  & 0.469  & 0.405           & 1.974        \\
                            &                      & subset &    0.404 &	0.447 &	0.489 &	0.340 &	0.362 &	0.447 &	0.415         &	1.851         \\
\multirow{2}{*}{BS-Net-HA} &                       & full    & 0.521 & 0.438  & 0.552  & 0.385 & 0.438  & 0.776  & 0.518           & 2.214        \\
                            &                      & subset & 0.458 & 0.479  & 0.521  & 0.375 & 0.479  & 0.625  & 0.490           & 2.396       \\
* from scratch              &                      & subset & 0.375 & 0.646  & 0.604  & 0.458 & 0.542  & 0.479  & 0.517           & 2.188       \\
* fine-tuning               &                      & subset & 0.479 & 0.500  & 0.500  & 0.354 & 0.458  & 0.500  & \textbf{0.465}           & 2.000       \\ %\midrule
\bottomrule
\end{tabular}
\label{tab:public-test}
\end{table*}

Looking at results on the full dataset, we can derive that, even by directly applying the model trained on the Brixia COVID-19 dataset on a completely different dataset (and collected in a highly uncontrolled way), the network confirms the meaningfulness of the learning task and shows a fair robustness to work in different context even in an uncontrolled way. 
On the other hand, the skilled human observer confirms higher generalization capability, with a MAE of 0.405 on the full dataset. 
On the reduced subset, when retrained from scratch, the network is not able anymore to produce even the results obtained by the same model trained on the Brixia COVID-19 dataset: a clear evidence of the need to work with a large dataset and of the adequate capacity of our model. 
This is further confirmed by looking at the fine-tuning results, where the network reaches the best performance (MAE 0.465) by exploiting the already trained baseline with only a tiny amount of training data.

%% -------------------------------------
%% SUBSECTION 6.6 - Err. and Explainab.
%% -------------------------------------

\subsection{Explainability maps}
\label{res:explain}

\begin{figure*}[t]
    \centering
    \includegraphics[width=1.0\textwidth]{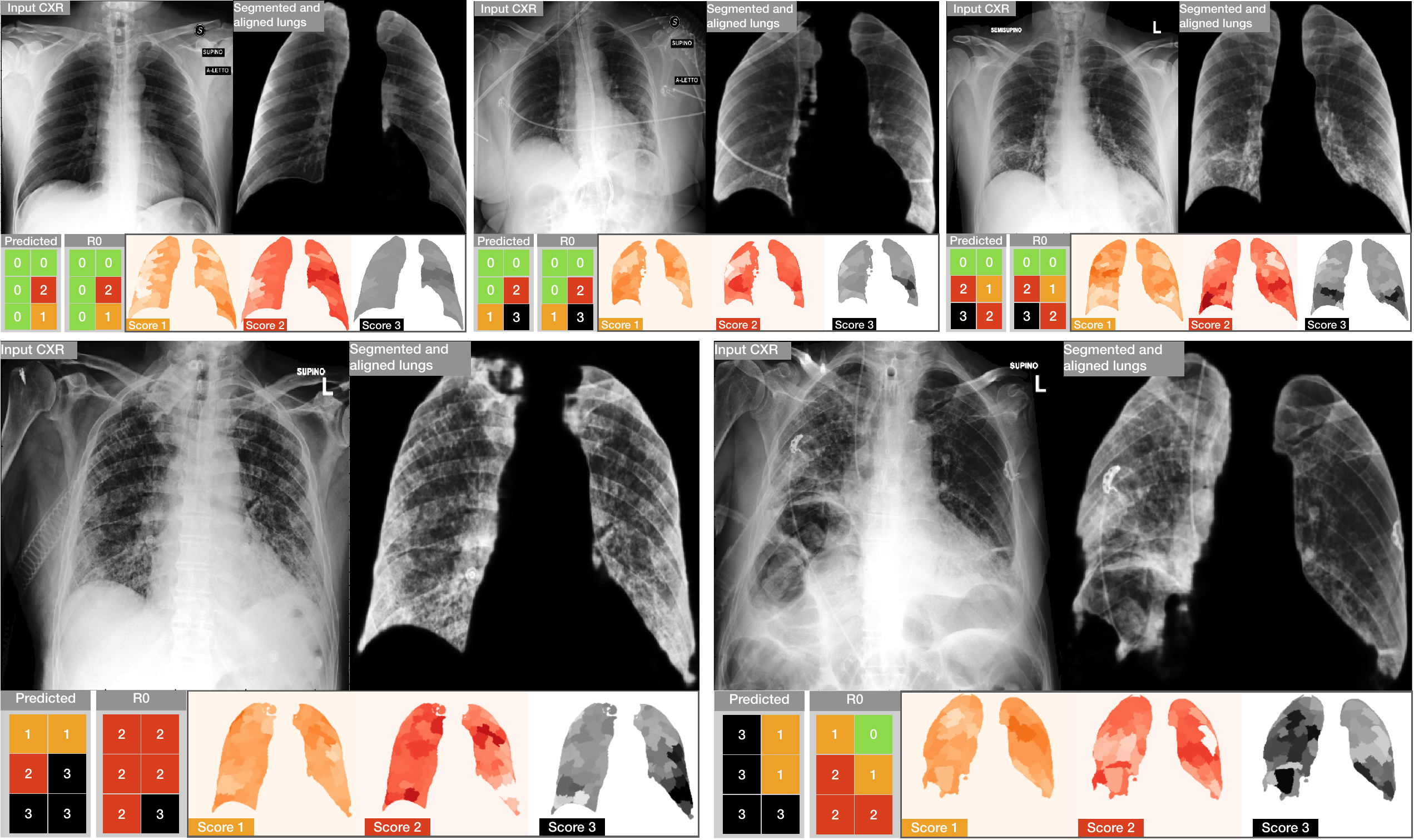}
    \caption{Results and related explainability maps obtained on five examples from the Brixia COVID-19 test set. (top) Three examples of accurate predictions. (bottom) Two critical cases in which the prediction is poor with respect to the original clinical annotation $R_0$. For each block, the most left image is the input CXR, followed by the aligned and masked lungs. In the second row we show the predicted \texttt{Brixia~score} with respect to the original clinical annotation $R_0$, and the explainability map. In such maps the relevance is colored so that white means that the region does not contribute to that prediction, while the class colour (i.e., 1 = orange, 2 = red, 3 = black) means that the region had an important role in the prediction of thaT~score class.}
    \label{fig:results}
\end{figure*}

In Figure~\ref{fig:results} we illustrate some explainability maps generated on Brixia COVID-19 data: three are exact predictions (top), while two are chosen among the most difficult cases (bottom).
Along with maps, we also report the related lung segmentation and alignment images.
Such maps, obtained as described in Section~\ref{sec:explain}, clearly highlight the regions that triggered a specific score: they are drawn with the colour of the class score (i.e., 1 = orange, 2 = red, 3 = black) in case they significantly contributed to the decision for that score, while a white coloured region means that it gave no contribution to such score decision.
The first row of Figure~\ref{fig:results} offers a clear overview of the agreement level between the network and the radiologist in shift ($R_0$).
Conversely, the two cases in the bottom row of Figure~\ref{fig:results} evidences both under- and over-estimations in single sectors, despite producing a correct Global Score in the case on the left.

%% -------------------------------------
%% SUBSECTION 6.5 - Ablation
%% -------------------------------------

\subsection{Ablation and variation studies}
\label{res:ablation}
We conducted various ablation and technology variation studies. First, we assess the actual contribution of the last training phase that involves all weights in the end-to-end configuration. Then, adopting BS-Net in the HA configuration, we conducted two sets of experiments regarding some modifications of feature extraction and data augmentation strategies, which are carried out on the training (3,313 CXRs) and on the validation (945 CXRs) sets of the Brixia COVID-19 database.
Finally, we carried out an experiment to evaluate the impact of the multi-resolution feature aligner.
\paragraph{End-to-end training}
With reference to the structured ``from-the-part-to-the-whole'' training strategy, the contribution of the last end-to-end training stage is significant, since it accounts for about 6-7\% of the MAE performance, for both Soft and Hard Attention configurations.

\paragraph{Feature Extraction}
This set of experiments investigates on the type and complexity of the feature map extraction leading to the \texttt{Brixia~score} estimation (see Scoring head in  Section~\ref{subsec:model}).
We then compare the adopted FPN-based solution with three different configurations.
The first, simplified, configuration gets rid of the multi-scale approach, so that the score estimation exploits the features extracted by the backbone head.
The second, more complex, configuration envisages an articulated multi-scale approach based on the EfficientDet \citep{tan2019efficientdet}, where BiFPN (Bidirectional FPN) blocks are introduced for an easy and fast multi-scale feature fusion.
The third configuration adds one resolution level to the FPN to allow the flow of $1,024\times1,024$ images. 
Results on this benchmark are reported in Table~\ref{tab:abl-architecture}.
While the first simplified configuration produces an improved MAE on the training set, we observe a poorer generalization capability on the validation set. 
The complex configuration, instead, produces a worsening on both data sets. 
Therefore FPN confirms to be a good intermediate solution between the latter two configurations.
Eventually, adding one resolution level produces again a performance worsening (this is a confident indication despite, for memory limitations, we did not succeeded in performing a complete fine-tuning).

\begin{table}[ht]
\centering
\small
\caption{Performance on the training and validation set for different Feature Pyramid Networks (or lack of).\\ $^{(*)}$ No complete fine-tuning due to memory limitation}
\begin{tabular}{@{}lll@{}}
\toprule
                & Train MAE (SD) & Val MAE (SD)\\ \midrule
FPN (adopted)            &    0.455 (0.578)       &     \textbf{0.469} (0.583)      \\
Backbone head   &    \textit{\textbf{0.395}}  (0.542)      &     0.475  (0.592)        \\
BiFPN           &    0.486 (0.593)       &    0.504 (0.598)           \\ 
FPN 1024 $^{(*)}$        &    0.498 (0.564)      &     0.498 (0.589)         \\
\bottomrule
\end{tabular}
\label{tab:abl-architecture}
\end{table}

\paragraph{Data Augmentation} The investigated matter regards whether the pre-processing and augmentation policies we adopted are effective, or if there are some prevailing or redundant constituents. 
In Table~\ref{tab:abl-augment} we present the results of performed experiments which combines different augmentation policies: it is clearly evident that partial augmentation policies are not adequate, since they produce performance worsening on the validation set, and tend to create overfitting gaps between training and validation performance.
Conversely, their joint use produces best results and not such a gap.
Moreover, we can appreciate the impact of the pre-processing equalization, which is able to correctly handle the grayscale variability in the dataset.

\begin{table}[ht]
\centering
\small
\caption{Performance on training and valid. sets for different augmentation policies.}
\begin{tabular}{@{}llll@{}}
\toprule             & Pre-proc.   & Train MAE (SD) & Val MAE (SD) \\ \midrule
No augmentation         & y &   0.306 (0.490)        &   0.528 (0.610)     \\
Bright. \& contrast  & y &    0.341 (0.518)       &   0.550 (0.628)     \\
Geometric transf.    & y &   0.272 (0.460)        &   0.541 (0.623)     \\
All together      & n &   0.437 (0.585)       &   0.571 (0.645)     \\
All together    & y &  0.455 (0.578)         &   0.469 (0.583)     \\
\bottomrule
\end{tabular}
\label{tab:abl-augment}
\end{table}

\paragraph{Multi-resolution feature aligner}
In order to demonstrate the robustness of the proposed network to misalignment and the contribution of the adopted multi-resolution feature alignment solution, we augmented the test set by synthetically rotate (from -30 to 30 degrees) all its images and we measured the network performance \textit{with} and \textit{without} the compensation estimated by the alignment block.

\begin{figure}
    \centering
    \includegraphics[width=0.47\textwidth]{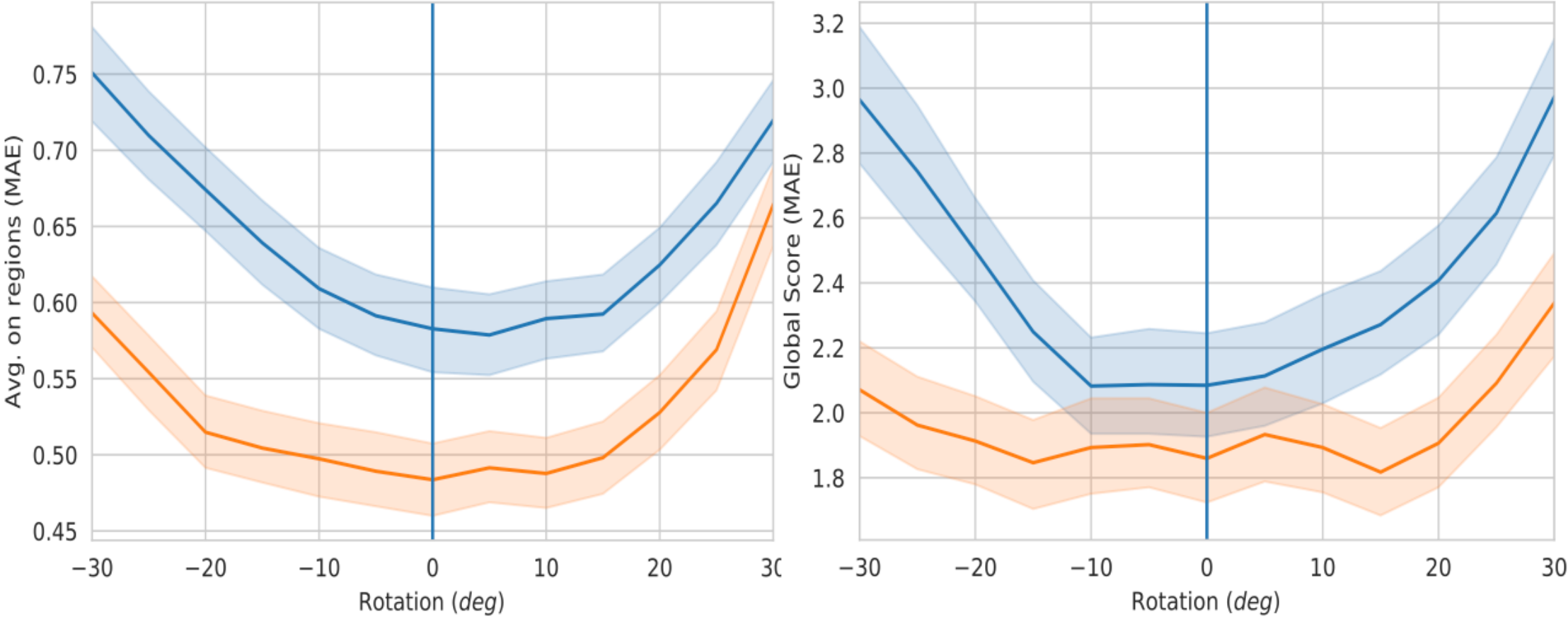}
    \caption{MAE on regions (left) and MAE of the global score (right) versus synthetic rotation. The blue curve is from the network 'without' the alignment block, while the orange is 'with' the alignment block enabled. The shaded areas correspond to the 95\% confidence interval.}
    \label{fig:rot-test}
\end{figure}

Figure \ref{fig:rot-test} shows how the MAE (on single regions, as well as globally) varies according to the value of the induced rotation angle.  
Looking at this picture, two important observations arise. First, even without induced rotations (vertical blue line at 0 degrees), the use of the alignment block produces significantly better results, and this demonstrates the effectiveness of the compensation of the real image misalignments originally present in the dataset. Second, the alignment module is able to compensate and improve the performance in a wide range of induced rotations, while without alignment, the influence of rotation angle on performance degradation is evident. 
Moreover, the flattened error region, approximately corresponding to the range $[-20^{\circ}, 20^{\circ}]$, is in line with the +/- 25 degree range used for the alignment block pre-training (Table \ref{tab:alignment-dataset}), and compatible with the actual range of rotations compensated by the alignment block during testing (we measured they span over a min/max range of +/- 15 degrees). Overall, the obtained performance improvement on MAE is about 20-25\% with respect to the case without alignment stage.
It is therefore clear from these results that the alignment network actually learns to compensate misalignment and its exploitation for multi-resolution feature alignment produces a significant performance boost.

\section{Discussion}
\label{sec:disc}

We have introduced an end-to-end image analysis system for the assessment of a semi-quantitative rating based on a highly difficult visual task. 
The estimated lung severity score is, by itself, the result of a compromise: on the one hand, the need for a clinically expressive granularity of the different stages of the disease; on the other hand, the built-in subjectivity in the interpretation of CXR images, stemming from the intrinsic limits of such imaging modality, and from the high variability of COVID-19 manifestations. %(also combined with variable acquisition conditions and parameters).
As an additional complication, the available \texttt{Brixia~score}, even if coming from expert personnel, has neither ground truth characteristics (as ratings are affected by inter-observer variability), nor it is highly accurate in terms of spatial indication (since scores are related to generic rectangular regions).

The best performance in the prediction of the \texttt{Brixia~score} is obtained, in all tests, by using an ensemble decision strategy (ENS). 
Reported mean absolute errors on the six regions are: MAE=0.441, when compared against the clinical annotations by radiologist $R_0$ on the whole test set of 450 CXRs; and MAE=0.424 when compared against a Consensus-based Gold Standard set of 150 images annotated by 5 radiologists.
%discussion starts here
Naively speaking, being the scoring scale defined on integers, any MAE measured on single regions below 0.5 could be interpreted as acceptable. 
This might appear as a simplified reasoning, but interviewed radiologists, with hundreds of cases of experience of such semi-quantitative rating system, also indicated, from a clinical perspective, $\pm 0.5$ as an acceptable error on each region of the  \texttt{Brixia~score}, and $\pm 2$ as an acceptable error on the Global Score from 0 to 18.
These indications are also backed up by the prognostic value and associated use of the score as a severity indicator that comes from the experimental evidence and clinical observations during the first period of its application \citep{borghesi2020third}. 

The above \textit{a-priori} interpretation of ``acceptable'' error on a single region is clearly not sufficient.
It is also for this reason that we built a CbGS. This is useful to have a reference measure of the inter-rater agreement between human annotators, acting as a boundary measure of human performance.
This is relevant since, being a source of error in our weakly supervised approach, such inter-rater agreement also determines an implicit limit to the performance we can expect from the network.
Tests on the CbGS confirm that, on average, the level of agreement reached between the network and any other human rater is slightly above the inter-rater agreements between any pair of human raters, thus statistically evidencing that BS-Net performance overcomes radiologists' in accomplishing the task. This is a fundamental basis to think at  clinically oriented studies in a perspective of a responsible deployment of the technology in (human-machine collaborative or computer-aided diagnosis) clinical settings.
A MAE under 0.5 for both the network and radiologists is also an indirect evidence of the fact that the \texttt{Brixia~score} rating system (on four severity values) is a good trade-off between the two opposite needs of having a fine granularity of the rating scale, and a good inter-rater agreement. Moreover, what comes out from the comparison with other scoring systems is that the \texttt{Brixia~score} can be considered as a good super-set with respect to others. This is provided by either a good spatial granularity (over six regions), and a good sensibility (over four levels). 
%For what concerns the grading scale, even if a geographic score based on the extent of lung involvement by ground glass opacity or consolidation may be very objective, our proposal do not present high levels of variability.

The fact that the ensemble decision strategy combines the strengths of the soft and hard attention policies deserves some further elaboration. 
In fact, if on one hand, removing the context is fine to avoid possible bias on the decision, on the other hand, the context info can help when the segmentation is too restrictive. 
For example, retrocardiac consolidations that could be visible are removed from the lung segmentation and therefore their assessment is not allowed in the hard attention approach. 
This residual complementarity between the two options is exploited by the ensemble decision and this can explain the significant improvement. 
This also clearly justifies the more structured approach we designed, which demonstrates to pay-off right from the soft-attention configuration.
Another aspect that emerges looking at MEr is that our architecture does not over/under-estimate, while the other compared architecture (ResNet-18) tends to overestimate.
A last qualitative evidence in favor of the use of the composite loss function of Equation ~\ref{eq-loss}, comprising an additional component related to the MAE, is that the performance we obtain are unbiased and stable over several repetitions of the whole training process.

Although aware that collecting a multi-center database would allow further generalizability tests and possibly increase the system robustness and portability, our clinical dataset is already highly representative of a large set of variability parameters: therefore, we do not expect to be varying too much between centers in terms of pneumonia severity range and distribution, adverse image acquisition conditions (ICU devices and patient state conditioned positioning), acquisition devices (fixed vs portable X-Ray equipment, CR vs DX, different manufacturers), patient age. Moreover, despite the geography of lungs does not depend on human phenotype, our patient sample also reflects the multi-ethnic composition of North-Italy population.
Portability tests performed on public datasets can be easily turned into clear guidelines for the use of the proposed model on data originated at different facilities and/or on related clinical contexts.
In fact, from collected evidence, there are all the makings of a highly portable model: performance are robust to change of settings within the COVID scenario, while transfer learning on quality data is highly advisable for performance optimization in different domains. Moreover, the results of this study suggests the possibility of using our system for monitoring the severity of potentially any pathology that manifests itself with a similar 'basic syntax' in terms of opacification of the parenchyma. However, the \textit{Brixia score} has been designed specifically for COVID severity assessment, and the only possible direct application of our model is the one concerning other viral pneumonia that lead to a ARDS (adult respiratory distress syndrome) clinical picture, for example that of the H1N1 virus or the one of other coronaviruses. For bacterial / mycobacterial, fungal, and other viral infections, or in case of pulmonary edema (where, unlike COVID pneumonia, pleural effusion dominates), other scoring might be preferable and domain adaptation would be needed.

For what concerns explainability maps, this information can be used to increase trust in the outcome, and point in possible confounding situations. 
With the proposed super-pixel-based approach we pave the way to new kinds of representations. In particular, we used a three-map representation to show the accurate localization power and the richness of generated information. In the following we account for both specific examples and more general considerations coming from our radiologists about the nature and the use of the proposed explainability solutions, as collected during CXR interpretation activities.
Looking for example at Figure~\ref{fig:results}, the case on the bottom-left part, despite producing a correct Global Score (considered a positive fact by the radiologists, given the specific case), evidences both under- and over-estimations in single sectors. 
Reviewed by a senior radiologist, this case evidenced some problematic areas even for the radiologists that are often called to express an average score about what they see in different areas. 
For this reason the super-pixel explainability maps constitute a useful machine-human communication tool. 
In particular the sector A interested by the presence of an assisted ventilation device can determine a common problem for both machine an expert, but the machine proves to not diverge toward unduly high ranking, while the expert ruled in favor to the machine in the other upper lung sector D, while confirms a slight overestimation in sector E. 
In general, we can appreciate the fact that external equipment do not harm the prediction, nor produces unwanted biases as in Figure~\ref{fig:results}~(top-middle, bottom-left, bottom-right). 
The last case (bottom right) evidences a segmentation error on the right lung, since part of the colon (filled by air, pulling up the diaphragm and compressing the lung) enters in the segmentation mask, determining a wrong evaluation  of the corresponding area. 
However, the radiologist positively considered the fact that the system sees a difference of two levels between the upper lung portions, which the expert considered a more coherent judgment at a second view. From these qualitative observations and explainability map support, we can conclude that, despite the automated AI-driven scoring is not meant to eliminate radiologists' evaluations, it can be used to aid and streamline the reporting workflow, and improving the timeliness of first evaluation, by proposing a preliminary interpretation of findings.

Starting from mid December 2020 the proposed system is undergoing an experimental deployment in the two Diagnostic Radiology Units of ASST Spedali Civili of Brescia (Italy), where all radiologists can see the severity score estimation and explainability maps during the reporting of all CXRs from incoming (admission/ER triage) and hospitalized COVID-19 patients.
In general, we had a valuable positive general feedback from radiologists that considered these maps as a possible complementary source of visual attention and of second level reasoning in a possible scenario of computer-assisted diagnosis.
%Finding the best representation with a deeper understanding of its utility in clinical practice is something that deserves dedicated experiments we would like to convey in a dedicated work. 
In the following we quote the opinion of a senior radiologist in the team, when interviewed about the proposed explainability approach:
``the maps concentrate in a small space a huge amount of information; in the overburdened workflow of a radiology department (not only during a pandemic, but also in the routine practice) it is quite unlikely that an experienced physician interrogates such a map. The reporting time of a CXR is in the range of a few seconds, thus the assessment of the explainability maps in every single case would significantly delay the reporting process; on the other hand, the maps are far from being useless. First of all, the neural network and the radiologist examine the same object (a CXR) using completely different perspectives and approaches; the map may be conceived as a mediator that allows subjects that speak different languages to communicate. In addition, in selected and complex cases (particularly during a multidisciplinary discussion) the maps may support the decision offering a valuable and non-subjective second opinion. Finally, in an academic setting the maps may guide young radiologists in identifying the abnormalities that sum up to a final diagnosis and to a definition of severity." Thus, even if explainability maps can be sometimes complex to interpret, finding good representations able to effectively convey visual information and to act as a mediator between clinicians and the AI model is of recognized high relevance. Combined with the fact that explainability solutions have an important role in emerging regulatory framework for trustable AI, it is our intention to ground on interesting properties of our super-pixel-based maps to further explore, in a dedicated work, their optimization and adoption according to user experience in clinical practice.
%Such aid might also prove useful in case of shortage of radiologists, condition that might be worsened due to the spread of contagion among physicians. 

The main limitations of this work are related to residual errors. 
Despite highly aligned with radiologists' performance and not evidencing statistical biases, a case-by-case deeper comparison of possible causes of error could guide further improvement of the image analysis architecture. 
In particular, since the segmentation task is sometimes ambiguous and hard to accomplish, better solutions should be designed to handle anomalies in the segmentation results.

%% -----------------------------------------
%%          SECTION 7 - FINAL REMARKS
%% -----------------------------------------

\section{Conclusions and future directions}
\label{sec:concl}

%--->DA INTRO A conclusion?
%Il nostro lavoro è il primo a proporre una soluzione originale e mirata al task spevifico. Inoltre la nostra stima è multiparametrica in quanto andamo a stmaree non solo lo score complessivo ma lo score su ciascuna area polminare.

Our study for the estimation of a semi-quantitative rating of lung severity is justified and driven by the strong clinical reasons related to the role of CXR images for the management of COVID-19 patient monitoring, especially in conditions of overloaded healthcare facilities. 
Working with a very large dataset of almost 5,000 images allowed to develop a solution which deals with all aspects of the problem, also thanks to other datasets which are exploited in a hierarchical end-to-end training process.
The prospected solution is designed to work in a weakly supervised context and to be capable of manifesting self-attentive abilities on data directly coming from all CXR modalities/devices and in all clinical and patient care conditions. 
Having tested architectural variants, and targeted ablation studies, the network performance ultimately surpasses qualified radiologists in rating accuracy and consistency.
We also collected evidences of the robustness and portability of the model to other clinical settings. 
%All annotated CXRs and the proposed models are made available for research purposes\footnote{Note for reviewers: the dataset is subject of a separate data-only journal submission while models will be released upon acceptance of this paper.}.

Regarding future work, observed results constitute a strong basis in a perspective of more clinically oriented validations and in a perspective of trustable and responsible deployment. An interesting clinical scenario to evaluate is the following: whether, in case the BS-Net follows the same patient, it could exhibits greater self-coherent behavior if compared to the case where serial CXR acquisitions are reported by different radiologists (according to availability and working shifts). 
Another direction to consider is the possibility to exploit CT images from COVID-19 patients to derive semi-quantitative ground truth information directly from the quantitative volumetric assessments.
\section*{Source code and Data availability}
All annotated CXRs and the proposed system (both source code and pre-trained models) are made available for research purposes at the BrixIA project page  \url{https://brixia.github.io}.

\section*{Declaration of Competing Interest}
None.

\section*{Acknowledgements}

The authors would like to thank colleagues of the Information and IT System Operational Units of ASST Spedali Civili di Brescia, Francesco Scazzoli, Silvio Finardi, Roberto Marini, and Sabrina Vicari for providing the infrastructure. A very special thanks to Marco Renato Roberto Merli, and Andrea Carola of Philips S.p.A., and to Fabio Capra, Daniele Soldi, Gian Stefano Bosio, and Simone Gaggero of EL.CO. S.r.l (Cairo Montenotte, Italy) for their outstanding technical support.

%\section*{Author Contributions}
%xxx

\bibliographystyle{model2-names.bst}\biboptions{authoryear}
\bibliography{bibliography}

\end{document}